\documentclass[MSNbibl,number,citesort,dvips]{arxbj}
\usepackage{upgreek,dcolumn}
\usepackage{graphicx}

\aid{0}
\volume{19}
\issue{3}
\pubyear{2013}
\firstpage{803}
\lastpage{845}
\doi{10.3150/12-BEJ476} 

\makeatletter
\newcolumntype{d}[1]{D{.}{.}{#1}}
\renewcommand{\pi}{\uppi}

\newcommand{\rrvert}{\vert}
\newcommand{\llvert}{\vert}
\newcommand{\eqref}[1]{(\ref{#1})}
\newtheorem{prop}{Proposition}
\newtheorem{cor}{Corollary}
\newremark{ex}{Example}
\newremark{rem}{Remark}

\newcommand{\R}{\mathbb{R}}
\newcommand{\E}{\mathbb{E}}

\newcommand{\ga}{\mathfrak{g}}

\def\N{\mathbb{N}}

\newcommand{\newnu}{\overline{\nu}}
\newcommand{\newlambda}{\overline{\lambda}}

\makeatother

\begin{document}
\begin{frontmatter}

\title{Modelling energy spot prices by volatility modulated L\'{e}vy-driven Volterra processes}
\runtitle{Modelling energy spot prices by LSS processes}

\begin{aug}
\author[1]{\fnms{Ole~E.}~\snm{Barndorff-Nielsen}\thanksref{1}\ead[label=e1]{oebn@imf.au.dk}},
\author[2]{\fnms{Fred~Espen}~\snm{Benth}\thanksref{2}\ead[label=e2]{fredb@math.uio.no}}
\and\break
\author[3]{\fnms{Almut~E.~D.}~\snm{Veraart}\thanksref{3}\ead[label=e3]{a.veraart@imperial.ac.uk}\corref{}}
\runauthor{O.E. Barndorff-Nielsen, F.E. Benth and A.E.D. Veraart} 
\address[1]{Thiele Center,
Department of Mathematical Sciences \& CREATES,
Department of Economics and Business,
Aarhus University,
Ny Munkegade 118,
DK-8000 Aarhus C,
Denmark
and
Technische Universit\"at M\"unchen,
Institute for Advanced Study,
85748 Garching,
Germany.\\
\printead{e1}}
\address[2]{Centre of Mathematics for Applications,
University of Oslo,
P.O. Box 1053, Blindern,
N--0316 Oslo,
Norway.
\printead{e2}}
\address[3]{Department of Mathematics,
Imperial College London,
180 Queen's Gate,
SW7 2AZ London,
United Kingdom
and
Creates.
\printead{e3}}
\end{aug}

\received{\smonth{3} \syear{2010}}
\revised{\smonth{4} \syear{2012}}

%
\begin{abstract}
This paper introduces the class of \textit{volatility modulated L\'
{e}vy-driven Volterra} ($\mathcal{VMLV}$)
processes and their important subclass of
\emph{L\'{e}vy semistationary} ($\mathcal{LSS}$) processes as a new
framework for modelling energy spot prices.
The main modelling idea consists of four principles: First,
deseasonalised spot prices can be modelled directly in stationarity.
Second, stochastic volatility is regarded as a key factor for modelling
energy spot prices. Third, the model allows for the possibility of
jumps and extreme spikes and, lastly, it features great flexibility in
terms of modelling the autocorrelation structure and the Samuelson effect.
We provide a detailed analysis of the probabilistic properties of
$\mathcal{VMLV}$ processes and show how they can capture many stylised
facts of energy markets.
Further, we derive forward prices based on our new spot price models
and discuss option pricing.
An empirical example based on electricity spot prices from the European
Energy Exchange confirms the practical relevance of our new modelling framework.

\end{abstract}

%
\begin{keyword}
\kwd{energy markets}
\kwd{forward price}
\kwd{generalised hyperbolic distribution}
\kwd{L\'{e}vy semistationary process}
\kwd{Samuelson effect}
\kwd{spot price}
\kwd{stochastic integration}
\kwd{stochastic volatility}
\kwd{volatility modulated L\'{e}vy-driven Volterra process}
\end{keyword}

\end{frontmatter}

\section{Introduction}\label{sec1}
Energy markets have been liberalised worldwide in the last two decades.
Since then we have witnessed the increasing importance of such
commodity markets which
organise the trade and supply of energy such as electricity, oil, gas
and coal. Closely related markets include also temperature and carbon markets.
There is no doubt that such markets will play a vital role in the
future given that the global demand for energy is constantly increasing.
The main products traded on energy markets are spot prices, futures and
forward contracts and options written on them.
Recently, there has been an increasing
research interest in the question of how such energy prices can be
modelled mathematically.
In this paper, we will focus on modelling energy \emph{spot} prices,
which include day-ahead as well as real-time prices.

Traditional spot price models typically allow for \emph{mean-reversion}
to reflect the fact that spot prices are determined as equilibrium
prices between supply and demand. In particular, they are commonly
based on a Gaussian Ornstein--Uhlenbeck (OU) process,
see Schwartz~\cite{Schwartz1997},
or more generally,
on weighted sums of OU processes with different levels of mean-reversion,
see, for example, Benth, Kallsen and
Meyer-Brandis~\cite{BenthKallsenMeyerBrandis2007} and Kl{\"u}ppelberg,
Meyer-Brandis and
Schmidt~\cite{KMBS2010}.
In such a modelling framework, the mean-reversion is modelled directly
or physically, by claiming that the price change is (negatively)
proportional to the current price.
In this paper, we interpret the mean-reversion often found in commodity
markets in a \emph{weak sense}
meaning that prices typically concentrate around a mean-level for
demand and supply reasons.
In order to account for such a weak form
mean-reversion, we suggest to use a modelling framework which allows
to model spot prices
(after seasonal adjustment) directly in \emph{stationarity}.
This paper proposes to use the class of volatility modulated L\'
{e}vy-driven Volterra ($\mathcal{VMLV}$) processes as the building
block for energy spot price models. In particular, the subclass of
so-called L\'{e}vy semistationary ($\mathcal{LSS}$) processes turns out
to be of high practical relevance.
Our main innovation lies in the fact that we
propose a modelling framework for energy spot prices which
(1) allows to model deseasonalised energy spot prices directly in \emph
{stationarity},
(2) comprises \emph{stochastic volatility},
(3) accounts for the possibility of \emph{jumps} and \emph{spikes},
(4) features great flexibility in terms of modelling the \emph
{autocorrelation structure} of spot prices and of describing the
so-called \emph{Samuelson effect}, which refers to the finding that the
volatility of a forward contract typically increases towards maturity.

We show that the new class of $\mathcal{VMLV}$ processes is
analytically tractable, and we will give a detailed account of the
theoretical properties of such processes.
Furthermore, we derive explicit expressions for the forward prices
implied by our new spot price model.
In addition, we will see that our new modelling framework
encompasses many classical models
such as those based on the Schwartz one-factor mean-reversion model,
see Schwartz~\cite{Schwartz1997}, and the wider class of continuous-time
autoregressive moving-average (CARMA) processes.
In that sense, it can also be regarded as a unifying modelling approach
for the most commonly used models for energy spot prices.
However,
the class of $\mathcal{VMLV}$ processes is much wider and directly
allows to model the key special features of energy spot prices and, in
particular, the stochastic volatility component.

The remaining part of the paper is structured as follows.
We start by introducing the class of $\mathcal{VMLV}$ processes in
Section~\ref{SectionLSS}.
Next, we formulate both a geometric and an arithmetic spot price model
class in
Section~\ref{SectionModel} and describe how our new models embed many
of the traditional models used in the recent literature.
In Section~\ref{SectionForward}, we derive the forward price dynamics
of the models and consider questions like
affinity of the forward price with respect to the underlying spot.
Section~\ref{Emp} contains an empirical example, where we study
electricity spot prices from the European Energy Exchange (EEX).
Finally, Section~\ref{SectionConclusion} concludes, and the \hyperref[SectionProofs]{Appendix}
contains the proofs of the main results.

\section{Preliminaries}
\label{SectionLSS}
Throughout this paper, we suppose that we have given a probability
space $(\Omega,\mathcal{F},P)$ with a filtration
$\mathfrak{F}=\{\mathcal{F}_t\}_{t\in\mathbb{R}}$ satisfying the
`usual conditions,' see
Karatzas and Shreve~\cite{KS}, Definition~I.2.25.
\subsection{The driving L\'{e}vy process}
Let $L=(L_t)_{t \geq0}$ denote a c\`{a}dl\`{a}g L\'{e}vy process with
L\'{e}vy--Khinchine representation
$\mathbb{E}(\exp(\mathrm{i} \zeta L_t))=\exp(t \psi(\zeta))$ for $t \geq0$,
$\zeta\in\mathbb{R}$ and
\begin{eqnarray*}
\psi(\zeta) = \mathrm{i}{d} \zeta- \frac{1}{2}\zeta^2 b+ \int
_{\mathbb
{R}} \bigl(\mathrm{e}^{\mathrm{i}\zeta z}- 1 - \mathrm{i}\zeta z
\mathbb{I}_{\{|z|\leq1\}} \bigr)\ell_L(\mathrm{d}z)
\end{eqnarray*}
for $d \in\mathbb{R}$, $b \geq0$ and the L\'{e}vy measure $\ell_L$
satisfying $\ell_L(\{0\})=0$ and $\int_{\mathbb{R}}(z^2\wedge1)\ell_L(\mathrm{d}z)< \infty$.
We denote the corresponding
characteristic triplet by $(d, b, \ell_L)$.
In a next step, we extend the definition of the L\'{e}vy process to a
process defined on the entire real line, by taking an independent copy
of $(L_t)_{t \geq0}$, which we denote by $(L^{*}_t)_{t \geq0}$ and we
define $L(t):= -L^*(-(t-))$ for $t < 0$. Throughout the paper
$L=(L_t)_{t\in\mathbb{R}}$ denotes such a two-sided L\'{e}vy process.
\subsection{Volatility modulated L\'{e}vy-driven Volterra processes}
The class of
volatility modulated L\'evy-driven Volterra ($\mathcal{VMLV}$)
processes, introduced by Barndorff-Nielsen and
Schmiegel~\cite{BNSchmiegel2008}, has the form
%
\begin{equation}
\label{eqsimpleambit} \overline{Y}_t=\mu+ \int_{-\infty}^tG(t,s)
\omega_{s-} \,\mathrm{d}L_s + \int_{-\infty}^tQ(t,s)a_s
\,\mathrm{d}s, \qquad t \in\mathbb{R},
\end{equation}
where $\mu$ is a constant, $L$ is the two-sided L\'{e}vy process
defined above, $G, Q\dvtx \mathbb{R}^2 \mapsto\mathbb{R}$ are
measurable deterministic functions with $G (t, s )
=Q (t, s ) =0$ for $t<s$, and $\omega=(\omega_t)_{t \in
\mathbb
{R}}$ and $a=(a_t)_{t \in\mathbb{R}}$ are c\`{a}dl\`{a}%
g stochastic processes which are (throughout the paper) assumed to be
\emph{independent} of $L$. In addition, we assume that $\omega$ is
positive. Note that such a process generalises the class of convoluted
subordinators defined in
Bender and
Marquardt~\cite{BenderMarquardt2009} to allow for stochastic volatility.

A very important subclass of $\mathcal{VMLV}$ processes is the new class
of L\'{e}vy semistationary ($\mathcal{LSS}$) processes: We choose two
functions $g, q\dvtx \mathbb{R}\mapsto\mathbb{R}_+$ such that
$G(t,s)=g(t-s)$ and $Q(t, s)=q(t-s)$ with $g(t-s)= q(t-s) = 0$ whenever
$s > t$, then
an $\mathcal{LSS}$ process $Y= \{ Y_{t} \}_{t\in\mathbb
{R}}$ is given by
%
\begin{equation}
Y_{t}=\mu+\int_{-\infty}^{t}g(t-s)
\omega_{s-}\,\mathrm{d}L_{s}+\int_{-\infty
}^{t}q(t-s)a_{s}
\,\mathrm{d}s,\qquad  t \in\mathbb{R}. \label{LSS}
\end{equation}
Note that the name L\'{e}vy semistationary
processes has been derived from the fact that the process~$Y$ is
stationary as soon as $\omega$ and $a$ are stationary.
In the case that $L=B$ is a two-sided Brownian motion, we call such
processes Brownian semistationary ($\mathcal{BSS}$) processes, which
have recently been introduced by Barndorff-Nielsen and
Schmiegel~\cite{BNSch09} in the context of
modelling turbulence in physics.

The class of
$\mathcal{LSS}$ processes
can be considered as the natural analogue for (semi-) stationary
processes of
L\'{e}vy semimartingales ($\mathcal{LSM}$), given by
\[
\mu+ \int_{0}^{t}\omega_{s-}\,
\mathrm{d}L_{s}+\int_{0}^{t}a_{s}
\,\mathrm {d}s,\qquad t \geq0.
\]
%
%
\begin{rem*}
The class of $\mathcal{VMLV}$ processes can be embedded into the class
of ambit fields,
see Barndorf-Nilsen and Schmiegel~\cite{BNSch04,BNSch07a}, Barndorff-Nielsen, Benth and
Veraart~\cite{BNBV2009,BNBV2009Forward}.

Also, it is possible to define $\mathcal{VMLV}$ and $\mathcal{LSS}$
processes for \emph{singular} kernel functions $G$ and~$g$,
respectively; a function $G$ (or $g$) defined as above is said to be
singular if $G(t,t-)$ (or $g(0+)$) does not exist or is not finite.
\end{rem*}

\subsection{Integrability conditions}
In order to simplify the exposition, we will focus on the stochastic
integral in the definition of an $\mathcal{VMLV}$ (and of an $\mathcal
{LSS}$) process only. That is, throughout the rest of the paper, let
%
\begin{eqnarray}
\label{ShortNot} \overline{Y}_t = \int_{-\infty}^t
G(t,s)\omega_{s-}\,\mathrm{d}L_s,\qquad
Y_t = \int_{-\infty}^t g(t-s)
\omega_{s-}\,\mathrm{d}L_s, \qquad t \in\mathbb{R}.
\end{eqnarray}
In this paper, we use the stochastic integration concept described in
Basse-O'Connor, Graversen and
Pedersen~\cite{BassePedersen2010} where a stochastic integration
theory on
$\mathbb{R}$, rather than on compact intervals as in the classical
framework, is presented. Throughout the paper, we assume that the
filtration $\mathfrak{F}$ is such that $L$ is a L\'{e}vy process with
respect to $\mathfrak{F}$, see Basse-O'Connor, Graversen and
Pedersen~\cite{BassePedersen2010}, Section 4, for details.

Let $(d, b, \ell_L)$ denote the L\'{e}vy triplet of $L$
associated with a truncation function $h(z)={\bf1}_{\{|z|\leq1\}}$.
According to Basse-O'Connor, Graversen and
Pedersen~\cite{BassePedersen2010}, Corollary 4.1, for $t \in
\mathbb{R}$ the process $(\phi_t(s))_{s \leq t}$ with $\phi_t(s):=G(t,s)\omega_{s-}$ is integrable with respect to $L$ if and only
if $(\phi_t(s))_{s \leq t}$ is $\mathfrak{F}$-predictable and the
following conditions hold almost surely:
%
\begin{eqnarray}
\label{mart-int-cond} b \int_{-\infty}^t
\phi_t(s)^2 \,\mathrm{d}s &<& \infty,\nonumber\\
\int
_{-\infty}^t \int_{\mathbb{R}} \bigl( 1
\wedge \bigl\llvert \phi_t(s) z\bigr\rrvert^2 \bigr)
\ell_L(\mathrm{d}z) \,\mathrm{d}s &<& \infty,
\\
\int_{-\infty}^t \biggl\llvert \,\mathrm{d}
\phi_t(s)+ \int_{\mathbb{R}} \bigl(h\bigl(z
\phi_t(s)\bigr) - \phi_t(s) h(z) \bigr)
\ell_L(\mathrm{d}z) \biggr\rrvert \,\mathrm{d}s&< &\infty.\nonumber
\end{eqnarray}
When we plug in $G(t, s) = g(t-s)$, we immediately obtain the
corresponding integrability conditions for the $\mathcal{LSS}$ 
process.
%
\begin{ex}\label{Ex1}
In the case of a Gaussian Ornstein--Uhlenbeck process, that is,
when\break
$g(t-s) = \exp(-\alpha(t-s))$ for $\alpha> 0$ and $\omega\equiv1$,
then the integrability conditions above are clearly satisfied, since we have
\[
b \int_{-\infty}^t\exp\bigl(-2\alpha(t-s)\bigr) \,
\mathrm{d}s = \frac{1}{2\alpha} b < \infty.
\]
\end{ex}
%
\subsubsection{Square integrability}
For many financial applications, it is natural to restrict the
attention to models where the variance is finite, and we focus
therefore on L\'evy processes $L$ with finite second moment.
Note that the integrability conditions above do not ensure
square-integrability of $\overline Y_t$ even if $L$ has finite second
moment. But substitute the first condition in (\ref{mart-int-cond})
with the stronger condition
%
\begin{equation}
\label{mart-int-l2-cond} \int_{-\infty}^t \mathbb{E}
\bigl(\phi_t(s)^2\bigr) \,\mathrm{d}s= \int
_{-\infty}^tG^2(t, s)\mathbb{E}\bigl[
\omega_s^2\bigr] \,\mathrm{d}s<\infty,
\end{equation}
then
$\int_{-\infty}^t G(t, s)\omega_{s-} \,\mathrm{d}(L_s - \mathbb{E}(L_s))$ is
square integrable. Clearly, $\mathbb{E}[\omega_s^2]$ is constant in
case of stationarity.
For the Lebesgue integral part, we need
%
\begin{equation}
\label{leb-int-l2-cond} \mathbb{E} \biggl[ \biggl(\int_{-\infty}^tG(t,
s)\omega_{s} \,\mathrm{d}s \biggr)^2 \biggr]<\infty.
\end{equation}
According to the Cauchy--Schwarz inequality, we find
\[
\mathbb{E} \biggl[ \biggl(\int_{-\infty}^tG(t,s)
\omega_{s} \,\mathrm{d}s \biggr)^2 \biggr]\leq\int
_{-\infty}^t\bigl|G(t,s)\bigr|^{2a} \,\mathrm{d}s\int
_{-\infty
}^t\bigl|G(t,s)\bigr|^{2(1-a)} \mathbb{E}\bigl[
\omega_s^2\bigr] \,\mathrm{d}s
\]
for any constant $a\in(0,1)$. Thus, a sufficient condition for (\ref
{leb-int-l2-cond}) to hold is that there exists an $a\in(0,1)$ such that
\begin{eqnarray*}
\int_{-\infty}^t\bigl|G(t,s)\bigr|^{2a} \,
\mathrm{d}s < \infty, \qquad \int_{-\infty
}^t\bigl|G(t,s)\bigr|^{2(1-a)}
\mathbb{E}\bigl[\omega_s^2\bigr] \,\mathrm{d}s < \infty,
\end{eqnarray*}
which simplifies to
%
\begin{eqnarray}
\int_0^{\infty}g^{2a}(x) \,\mathrm{d}x <
\infty,\qquad \int_{-\infty}^tg^{2(1-a)}(t-s)
\mathbb{E}\bigl[\omega_s^2\bigr] \,\mathrm{d}s<\infty,
\end{eqnarray}
in the $\mathcal{LSS}$ case.
Given a model for $\omega$ and $g$, these conditions are simple to verify.
Let us consider an example.
%
\begin{ex}
In Example~\ref{Ex1}, we showed that
for
the kernel function $g(x)=\exp(-\alpha x)$
and in the case of constant volatility,
the conditions\vadjust{\goodbreak} (\ref{mart-int-cond}) are satisfied. Next, suppose that
there is stochastic volatility, which
is defined by the Barndorff-Nielsen and
Shephard~\cite{BNS2001a} stochastic volatility model, that is
$\omega_s^2 = \int_{-\infty}^s\mathrm{e}^{-\lambda(s-u)}\,\mathrm{d}U_{\lambda s}$, for $s
\in\mathbb{R}$, $\lambda> 0$ and
a subordinator $U$. Suppose now that $U$ has cumulant function
$\int_0^{\infty}(\exp(\mathrm{i}\theta z)-1) \ell_U(\mathrm{d}z)$ for a L\'evy
measure $\ell_U$ supported on the positive real axis, and that $U_1$
has finite expectation. In this case, we have that
$
\mathbb{E} [\omega_s^2 ]
=
\int_0^{\infty}z \ell_U(\mathrm{d}z)<\infty$ for all $s$.
Thus, both (\ref{mart-int-l2-cond}) and (\ref{leb-int-l2-cond}) are
satisfied (the latter can be seen after using the sufficient
conditions), and we find that $Y_t$ is a square-integrable stochastic process.
\end{ex}

\section{The new model class for energy spot prices}\label{SectionModel}
This section presents the new modelling framework for energy spot
prices, which is based on $\mathcal{VMLV}$ processes.
As before, for ease of exposition, we will disregard the drift part in
the general $\mathcal{VMLV}$ process for most of our analysis and
rather use
$\overline Y=(\overline Y_t)_{t\in\mathbb{R}}$ with
%
\begin{equation}
\label{DefYSpot} \overline Y_t = \int_{-\infty}^t
G(t, s) \omega_{s-} \,\mathrm{d}L_s
\end{equation}
as the building block for energy spot price, see (\ref{eqsimpleambit})
for the precise definition of all components.
Throughout the paper, we assume that the corresponding integrability
conditions hold.
We can use the
$\mathcal{VMLV}$ process defined in (\ref{DefYSpot}) as the building
block to define both a geometric and an arithmetic model for the energy
spot price. Also, we need to account for trends and seasonal effects.
Let $\Lambda\dvtx  [0, \infty)\to[0, \infty)$ denote a
bounded and measurable deterministic seasonality and trend function.

In a \emph{geometric} set up, we
define the spot price $S^g=(S^g_t)_{t \geq0}$ by
%
\begin{equation}
\label{eqgambit-spot} S_t^g=\Lambda(t)\exp(\overline
Y_t),\qquad  t \geq0.
\end{equation}
In such a modelling framework, the deseasonalised, logarithmic spot
price is given by a $\mathcal{VMLV}$ process.
Alternatively, one can construct a spot price model which is of \emph
{arithmetic} type.
In particular,
we define the electricity spot price $S^{a} = (S_t^{a})_{t \geq0}$ by
%
\begin{equation}
\label{eqaambit-spot} S_t^{a} = \Lambda(t) + \overline
Y_t, \qquad t \geq0.
\end{equation}
(Note that the seasonal function $\Lambda$ in the geometric and the
arithmetic model is typically not the same.)
For general asset price models, one usually formulates conditions
which ensure that prices can only take positive values.
We can easily ensure positivity of our arithmetic model by imposing
that $L$ is a L\'{e}vy subordinator and that the kernel function $G$
takes only positive values.

\subsection{Model properties}
\subsubsection{Possibility of modelling in stationarity}
We have formulated the new spot price model in the general form based
on a $\mathcal{VMLV}$ process to be able to account for non-stationary
effects, see, for example, Burger \textit{et~al.}~\cite{BKMS2003}, Burger,
Graeber and
Schindlmayr~\cite{BGS2007}.\vadjust{\goodbreak}
If the empirical data analysis, however, supports the assumption of
working under stationarity, then we will restrict ourselves to the
analysis of $\mathcal{LSS}$ processes with stationary stochastic volatility.
As mentioned in the \hyperref[sec1]{Introduction}, traditional models for energy spot
prices are typically based on mean-reverting stochastic processes, see,
for example, Schwartz~\cite{Schwartz1997}, since
such a modelling framework reflects the fact that commodity spot prices
are equilibrium prices determined by supply and demand. Stationarity
can be regarded as a weak form of mean-reversion and is often found in
empirical studies on energy spot prices; one such example will be
presented in this paper.
\subsubsection{The initial value}
In order to be able to have a stationary model,
the lower integration bound in the definition of the $\mathcal{VMLV}$
process, and in particular for the $\mathcal{LSS}$ process, is chosen
to be $-\infty$ rather than~0.
Clearly, in any real application, we observe data from a starting value
onwards, which is traditionally chosen as the observation at time $t=0$.
Hence,
while $\mathcal{VMLV}$ processes are defined on the entire real line,
we only define the spot price for $t\geq0$.
The observed initial value of the spot price at time $t=0$
is assumed to be a \textit{realisation} of
the random variable $S^{g}_0=\Lambda(0)\exp(\overline Y_0)$ and
$S_0^{a} = \Lambda(0)+\overline Y_0$, respectively.
Such a choice guarantees that the deseasonalised spot price is a
stationary process, provided we are in the stationary $\mathcal{LSS}$
framework.

\subsubsection{The driving L\'{e}vy process}
Since $\mathcal{VMLV}$ and
$\mathcal{LSS}$ processes are driven by
a general L\'{e}vy process $L$, it is possible to account for price
jumps and spikes, which are often observed in electricity markets. At
the same time, one can also allow for Brownian motion-driven models,
which are very common in, for example, temperature markets, see, for example, Benth,
H{\"a}rdle and
Cabrera~\cite{BenthCabreraHaerdle2009}.

\subsubsection{Stochastic volatility}
A key ingredient of our new modelling framework which sets the model
apart from many traditional models is the fact that it allows for
stochastic volatility.
Volatility clusters are often found in energy prices, see, for example,
Hikspoors and Jaimungal~\cite{HJ},
Trolle and
Schwartz~\cite{TrolleSchwartz2009},
Benth~\cite{B},
Benth and Vos~\cite{BenthVos},
Koopman, Ooms and
Carnero~\cite{Koopmanetal2007},
Veraart and Veraart~\cite{VV2}.
Therefore, it is important to have a stochastic volatility component,
given by $\omega$, in the model.
Note that a very general model for the volatility process would be to
choose an $\mathcal{VMLV}$ process, that is, $\omega^2_t=Z_t$ and
%
\begin{equation}
\label{eqambitsv} Z_t=\int_{-\infty}^ti(t,s)
\,\mathrm{d}U_s ,
\end{equation}
where $i$ denotes a deterministic, positive function and $U$ is a L\'
{e}vy subordinator. In fact, if we want to ensure that the volatility
$Z$ is stationary, we can work with a function of the form $i(t,s)=
i^*(t-s)$, for a deterministic, positive function $i^*$.

\subsubsection{Autocorrelation structure and Samuelson effect}
The kernel function $G$ (or $g$) plays a vital role in our model and introduces
a flexibility which many traditional models lack: We will see in
Section~\ref{SecEVarCov} that the kernel function -- together with the
autocorrelation function of the stochastic volatility process --
determines the autocorrelation function of the process $\overline Y$.
Hence our $\mathcal{VMLV}$ -- based models are able to produce various
types of autocorrelation functions depending on the choice of the
kernel function $G$.
It is important to stress here that this can be achieved by using \emph
{one} $\mathcal{VMLV}$ process only, whereas some traditional models
need to introduce a multi-factor structure to obtain a comparable
modelling flexibility.
Also due to the flexibility in the choice of the kernel function, we
can achieve greater flexibility in modelling the shape of the Samuelson
effect often observed in forward prices, including the hyperbolic one
suggested by Bjerksund, Rasmussen and
Stensland~\cite{BRS} as a reasonable volatility feature in power markets.
Note that we obtain the modelling flexibility in terms of the general
kernel function $G$ here since we specify our model directly through a
stochastic integral whereas most of the traditional models are
specified through evolutionary equations, which limit the choices of
kernel functions associated with solutions to such equations.
In that context, we note that a $\mathcal{VMLV}$ or an $\mathcal{LSS}$
process cannot in general be written in form of a stochastic
differential equation (due to the non-semimartingale character of the
process). In Section~\ref{SectSMCond}, we will discuss sufficient
conditions which ensure that an $\mathcal{LSS}$ process is a semimartingale.

\subsubsection{A unifying approach for traditional spot price
models}\label{SectUnifying}
As already mentioned above, energy spot prices are typically modelled
in stationarity, hence the class of $\mathcal{LSS}$ processes is
particularly relevant for applications.
In the following, we will show that many of the traditional spot price
models can be embedded into our
$\mathcal{LSS}$ process-based framework.

Our new framework nests the stationary version of the classical
one-factor Schwartz~\cite{Schwartz1997} model
studied for oil prices.
By letting $L$ be a L\'evy process with the pure-jump part given as a
compound Poisson process, Cartea and Figueroa~\cite{CF} successfully
fitted the Schwartz
model to
electricity spot prices in the UK market. Benth and {\v{S}}altyt{\.
e} Benth~\cite{BSB-energy} used a
normal inverse Gaussian L\'evy process $L$ to model UK spot gas and
Brent crude oil spot prices.
Another example which is nested by the class of $\mathcal{LSS}$
processes is a model
studied in Benth~\cite{B} in the context of gas markets, where the
deseasonalised logarithmic spot price dynamics is assumed to follow a
one-factor Schwartz process with stochastic volatility.
A more general class of models which is nested is the class of
so-called CARMA-processes,
which has been successfully used in temperature modelling and weather
derivatives pricing, see Benth, {\v{S}}altyt{\. e} Benth and
Koekebakker~\cite{BSBK-temp},
Benth, H{\"a}rdle and
L{\'o}pez Cabrera~\cite{BenthCabreraHaerdle2009} and H{\"a}rdle and
L{\'o}pez Cabrera~\cite{CabreraHaerdle2009}, and more
recently for electricity prices by
Garc{\'{\i}}a, Kl{\"u}ppelberg and
M{\"u}ller~\cite{GKM2010},
Benth \textit{et al.}~\cite{BenthKlueppelberVos}. A CARMA process is the
continuous-time analogue of an ARMA time series, see Brockwell \cite
{Brockwell2001a},
Brockwell~\cite{Brockwell2001b} for definition and details. More
precisely, suppose that for nonnegative integers $p>q$
\[
Y_t=\mathbf{b}'\mathbf{V}_t ,
\]
where $\mathbf{b}\in\R^p$ and $\mathbf{V}$ is a $p$-dimensional OU
process of the form
%
\begin{equation}
\label{multi-OU-carma} \mathrm{d}\mathbf{V}_t=\mathbf{A}\mathbf{V}_t
\,\mathrm{d}t+\mathbf{e}_p\,\mathrm{d}L_t,
\end{equation}
with
\[
\mathbf{A}=\left[
\matrix{ \mathrm{0} &
\mathbf{I}_{p-1}
\vspace*{2pt}\cr
-\alpha_p & -\alpha_{p-1}\cdots-\alpha_1}
\right] .
\]
Here we use the notation $\mathbf{I}_{p-1}$ for the $(p-1)\times
(p-1)$-identity matrix, $\mathbf{e}_p$ the $p$th coordinate vector
(where the first $p-1$ entries are zero and the $p$th entry is 1) and
$\mathbf{b}'=[b_0, b_{1},\ldots,b_{p-1}]$ is the transpose of
$\mathbf
{b}$, with $b_q=1$ and $b_j=0$ for $q<j<p$.
In Brockwell~\cite{Brock}, it is shown that if all the eigenvalues of
$\mathbf{A}$
have negative real parts, then
$(\mathbf{V}_t)_{t \in\mathbb{R}}$ defined as
\[
\mathbf{V}_t=\int_{-\infty}^t{
\mathrm{e}}^{\mathbf{A}(t-s)}\mathbf {e}_p \,\mathrm{d}L(s) ,
\]
is the (strictly) stationary solution of \eqref{multi-OU-carma}. Moreover,
%
\begin{equation}
\label{CARMA} Y_t= \mathbf{b}'\mathbf{V}_t
= \int_{-\infty}^t\mathbf{b}'{\mathrm
{e}}^{{\bf
A}(t-s)}\mathbf{e}_p \,\mathrm{d}L(s) ,
\end{equation}
is a $\operatorname{CARMA}(p, q)$ process.
Hence, specifying $g(x)=\mathbf{b}'\exp(\mathbf{A}x)\mathbf{e}_p$ in
(\ref
{CARMA}), the
log-spot price dynamics will be an $\mathcal{LSS}$ process, but without
stochastic volatility.
Garc{\'{\i}}a, Kl{\"u}ppelberg and
M{\"u}ller~\cite{GKM2010}
argue for $\operatorname{CARMA}(2,1)$ dynamics as an appropriate class of
models for the deseasonalised log-spot price at the Singapore New
Electricity Market. The innovation process
$L$ is chosen to be in the class of stable processes. From Benth, {\v
{S}}altyt{\. e} Benth and
Koekebakker~\cite{BSBK-temp}, Brownian motion-driven $\operatorname{CARMA}(3,0)$ models
seem appropriate for modelling daily average temperatures, and are
applied for temperature derivatives pricing, including forward
price dynamics of various contracts. More recently, the dynamics of
wind speeds have been modelled by a Brownian motion-driven $\operatorname{CARMA}(4,0)$
model, and applied to wind derivatives pricing, see Benth and {\v
{S}}altyt{\.
e} Benth~\cite{BSB-wind}
for more details.

Finally note that the arithmetic model based on a superposition of
$\mathcal{LSS}$ processes nests the non-Gaussian Ornstein--Uhlenbeck
model which has recently been proposed for modelling electricity spot
prices, see Benth, Kallsen and
Meyer-Brandis~\cite{BenthKallsenMeyerBrandis2007}.

We emphasis again that, beyond the fact that $\mathcal{LSS}$ processes
can be regarded as a unifying modelling approach which nest many of the
existing spot price models, they also open up for entirely \emph{new}
model specifications, including more general choices of the kernel
function (resulting in non-linear models) and the presence of
stochastic volatility.

\subsection{Second order structure}\label{SecEVarCov}
Next, we study the second order structure of volatility modulated
Volterra processes
$\overline Y = (\overline Y_t)_{t \in\mathbb{R}}$, where
$\overline Y_t = \int_{-\infty}^t G(t,s) \omega_{s-} \,\mathrm{d}L_s$,
assuming the integrability conditions
(\ref{mart-int-cond}) hold and that in addition $\overline Y$ is square\vadjust{\goodbreak}
integrable.
Let $\kappa_1 = \mathbb{E}(L_1)$ and
$\kappa_2 = \operatorname{Var}(L_1)$. Recall that throughout the paper we assume that
the stochastic volatility $\omega$ is independent of the driving L\'
{e}vy process. Note that proofs of the following results are easy and
hence omitted.
%
\begin{prop}
The conditional second order structure of $\overline Y$ is given by
\begin{eqnarray*}
\mathbb{E}(\overline Y_t |\omega) &=& \kappa_1 \int
_{-\infty}^t G(t,s) \omega_s \,
\mathrm{d}s,\qquad \operatorname{Var}(\overline Y_t|\omega)= \kappa_2 \int
_{-\infty}^t G(t,s)^2
\omega_s^2 \,\mathrm{d}s,
\\[-2pt]
\operatorname{Cov}\bigl((\overline Y_{t+h}, \overline Y_t)|\omega\bigr)&=&
\kappa_2 \int_{-\infty}^{t} G(t+h,s) G(t,s)
\omega_s^2\,\mathrm{d}s \qquad\mbox{for } t \in\mathbb{R}, h
\geq0.
\end{eqnarray*}
\end{prop}
%
%
\begin{cor}
The conditional second order structure of $Y$ is given by
\begin{eqnarray*}
\mathbb{E}(Y_t |\omega) &=& \kappa_1 \int
_0^{\infty} g(x) \omega_{t-x} \,
\mathrm{d}x,\qquad \operatorname{Var}(Y_t|\omega)= \kappa_2 \int
_0^{\infty} g(x)^2 \omega_{t-x}^2
\,\mathrm{d}x,
\\[-2pt]
\operatorname{Cov}\bigl((Y_{t+h}, Y_t)|\omega\bigr)&=& \kappa_2
\int_0^{\infty} g(x+h)g(x)\omega_{t-x}^2
\,\mathrm{d}x \qquad\mbox{for } t \in\mathbb{R}, h \geq0.
\end{eqnarray*}
\end{cor}
The unconditional second order structure of $\overline Y$ is then given
as follows.
%
\begin{prop}
The second order structure of $\overline Y$ for stationary $\omega$ is
given by
\begin{eqnarray*}
\mathbb{E}(\overline Y_t) &=& \kappa_1 \mathbb{E}(
\omega_0) \int_{-\infty
}^t G(t,s)\,
\mathrm{d}s,
\\[-2pt]
\operatorname{Var}(\overline Y_t) &=& \kappa_2 \mathbb{E} \bigl(
\omega_0^2 \bigr) \int_{-\infty}^t
G(t,s)^2 \,\mathrm{d}s + \kappa_1^2 \int
_{-\infty}^t \int_{-\infty}^t
G(t,s)G(t,u) \gamma \bigl(|s-u|\bigr) \,\mathrm{d}s \,\mathrm{d}u,
\\[-2pt]
\operatorname{Cov}(\overline Y_{t+h}, \overline Y_t) &=&
\kappa_2 \mathbb{E} \bigl(\omega_0^2 \bigr)
\int_{-\infty}^{t} G(t+h,s)G(t,s) \,\mathrm{d}s
\\[-2pt]
&&{} + \kappa_1^2 \int_{-\infty}^{t+h}
\int_{-\infty}^{t} G(t+h,s)G(t,u) \gamma\bigl(|s-u|\bigr) \,
\mathrm{d}s \,\mathrm{d}u,
\end{eqnarray*}
where $\gamma(h) = \operatorname{Cov}(\omega_{t+h}, \omega_t)$ denotes the
autocovariance function of $\omega$, for $t \in\mathbb{R}, h \geq0$.
\end{prop}

The unconditional second order structure of $Y$ is then given as follows.
%
\begin{cor}
The second order structure of $Y$ for stationary $\omega$ is given by
\begin{eqnarray*}
\mathbb{E}(Y_t) &=& \kappa_1 \mathbb{E}(
\omega_0) \int_0^{\infty} g(x)\,
\mathrm{d}x,
\\[-2pt]
\operatorname{Var}(Y_t) &=& \kappa_2 \mathbb{E} \bigl(
\omega_0^2 \bigr) \int_0^{\infty}
g(x)^2 \,\mathrm{d}x + \kappa_1^2 \int
_0^{\infty} \int_0^{\infty}
g(x)g(y) \gamma \bigl(|x-y|\bigr) \,\mathrm{d}x \,\mathrm{d}y,
\\[-2pt]
\operatorname{Cov}(Y_{t+h}, Y_t) &=& \kappa_2 \mathbb{E} \bigl(
\omega_0^2 \bigr) \int_{0}^{\infty}
g(x+h)g(x) \,\mathrm{d}x + \kappa_1^2 \int
_{0}^{\infty} \int_0^{\infty}
g(x+h)g(y) \gamma\bigl(|x-y|\bigr) \,\mathrm{d}x \,\mathrm{d}y,
\end{eqnarray*}
where $\gamma(x) = \operatorname{Cov}(\omega_{t+x}, \omega_t)$ denotes the
autocovariance function of $\omega$, for $t \in\mathbb{R}, h \geq0$.
Hence, we have
%
\begin{eqnarray}
\label{Cor} &&\operatorname{Cor}(Y_{t+h}, Y_t)
\nonumber
\\[-8pt]
\\[-8pt]
\nonumber
&&\quad=\frac{\kappa_2 \mathbb{E} (\omega_0^2
) \int_0^{\infty} g(x+h)g(x) \,\mathrm{d}x + \kappa_1^2 \int_0^{\infty} \int_0^{\infty
}g(x+h)g(y) \gamma(|x-y|)\,\mathrm{d}x \,\mathrm{d}y}{\kappa_2 \mathbb{E} (\omega_0^2 ) \int_0^{\infty} g(x)^2 \,\mathrm{d}x + \kappa_1^2 \int_0^{\infty
} \int_0^{\infty}g(x)g(y) \gamma(|x-y|)\,\mathrm{d}x \,\mathrm{d}y}.
\end{eqnarray}
\end{cor}

%
\begin{cor}
If $\kappa_1=0$ or if $\omega$ has zero autocorrelation, then
\[
\operatorname{Cor}(Y_{t+h}, Y_t)=\frac{ \int_0^{\infty} g(x+h)g(x) \,\mathrm{d}x }{ \int_0^{\infty} g(x)^2 \,\mathrm{d}x }.
\]
\end{cor}
The last corollary shows that we get the same autocorrelation function
as in the ${\mathcal BSS}$ model. From the results above, we clearly
see the influence of the general damping function $g$ on the
correlation structure. A particular choice of $g$, which is interesting
in the energy context is studied in the next example.
%
\begin{ex}
\label{ex-bjerksund-g}
Consider the case
$
g(x)=\frac{\sigma}{x+b},
$
for $\sigma, b > 0$ and $\omega\equiv1$,
which is motivated from the forward model of Bjerksund, Rasmussen and
Stensland~\cite{BRS}, which we shall
return to in Section~\ref{SectionForward}.
We have that
$
\int_0^{\infty}g^2(x) \,\mathrm{d}x=\frac{\sigma^2}{b} .
$
This ensures integrability of $g(t-s)$ over $(-\infty,t)$ with respect
to any square integrable martingale L\'evy process $L$.
Furthermore,
$\int_0^{\infty}g(x+h)g(x) \,\mathrm{d}x
=\frac{\sigma^2}{h}\ln (1+\frac{h}{b} )$.
Thus,
\[
\operatorname{Cor}(Y_{t+h},Y_t)=\frac{b}{h}\ln \biggl(1+
\frac{h}{b} \biggr) .
\]
Observe that since $g$ can be written as
\[
g(x)= \frac{\sigma}{x+b}= \int_0^x
\frac{-\sigma \,\mathrm{d}s}{(s+b)^2} + \frac
{\sigma}{b} ,
\]
it follows that the
process $Y(t)=\int_{-\infty}^tg(t-s) \,\mathrm{d}B_s$ is a semimartingale
according to the Knight condition, see Knight~\cite{Kni92} and also
Basse~\cite{Basse2008El},
Basse and Pedersen~\cite{BassePedersen2009},
Basse-O'Connor, Graversen and
Pedersen~\cite{BassePedersen2010}.
\end{ex}

\subsection{Semimartingale conditions and absence of arbitrage}\label
{SectSMCond}
We pointed out that the subclass of $\mathcal{LSS}$ processes are
particularly relevant for modelling energy spot prices since they allow
one to model directly in stationarity. Let us focus on this class in
more detail. Clearly, an $\mathcal{LSS}$ process is in general not a
semimartingale. However, we can formulate sufficient conditions on the
kernel function and on the stochastic volatility component which ensure
the semimartingale property. The sufficient conditions are in line with
the conditions formulated for $\mathcal{BSS}$ processes in
Barndorff-Nielsen and
Schmiegel~\cite{BNSch09}, see also Barndorff-Nielsen and
Basse-O'Connor~\cite{BNBasse2009}.
Note that the proofs of the following results are provided in the \hyperref[SectionProofs]{Appendix}.
%
\begin{prop}\label{ThmSM}
Let $Y$ be an $\mathcal{LSS}$ process as defined in (\ref{LSS}).
Suppose the following conditions hold:
\begin{longlist}[(iii)]
\item[(i)] $\mathbb{E}|L_1| < \infty$.
\item[(ii)] The function values $g(0+)$ and $q(0+)$ exist and are finite.
\item[(iii)] The kernel function $g$ is absolutely continuous with
square integrable derivative~$g'$.
\item[(iv)] The process $(g'(t-s)\omega_{s-})_{s\in\R}$ is square
integrable for each $t \in\R$.
\item[(v)] The process $(q'(t-s)a_{s})_{s\in\R}$ is integrable for
each $t \in\R$.
\end{longlist}
Then $(Y_t)_{t\geq0}$ is a semimartingale with representation
%
\begin{equation}
\label{SMRep} Y_t = Y_0 + g(0+) \int
_0^t\omega_{s-}\,\mathrm{d}\overline
L_s + \int_0^t A_s \,
\mathrm{d}s \qquad\mbox{for } t \geq0,
\end{equation}
where $\overline L_s = L_s -\mathbb{E}(L_s)$ for $s \in\mathbb{R}$ and
\begin{eqnarray*}
A_s = g(0+)\omega_{s-}\mathbb{E}(L_1)+\int
_{-\infty}^s g'(s-u)
\omega_{u-}\,\mathrm{d}L_u + q(0+)a_s + \int
_{-\infty}^sq'(s-u)a_u \,
\mathrm{d}u.
\end{eqnarray*}
\end{prop}

%
\begin{ex}
An example of a kernel function which satisfies the above conditions is
given by
\begin{eqnarray*}
g(x) = \sum_{i=1}^J w_i
\exp(-\lambda_i x)\qquad \mbox{for } \lambda_i >0,
w_i \geq0, i=1,\ldots, J.
\end{eqnarray*}
For $J=1$, $Y$ is given by a volatility modulated Ornstein--Uhlenbeck process.
\end{ex}
In a next step, we are now able to find a representation for the
quadratic variation of an $\mathcal{LSS}$ process provided the
conditions of Proposition~\ref{ThmSM} are satisfied.

%
\begin{prop}\label{ThmQV}
Let $Y$ be an $\mathcal{LSS}$ process and suppose that the sufficient
conditions for $Y$ to be a semimartingale (as formulated in Proposition
\ref{ThmSM}) hold. Then, the quadratic variation of $Y$ is given by
\[
[Y]_t = g(0+)^2 \int_0^t
\omega_{s-}^2\,\mathrm{d}[L]_s\qquad \mbox{for } t
\geq0.
\]
\end{prop}
Note that the quadratic variation is a prominent measure of accumulated
stochastic volatility or intermittency over a certain period of time
and, hence, is a key object of interest in many areas of application
and, in particular, in finance.

The question of deriving semimartingale conditions for $\mathcal{LSS}$
processes is closely linked to the question whether a
spot price model based on an $\mathcal{LSS}$ process is prone to
arbitrage opportunities.
In classical financial theory, we usually stick to the semimartingale
framework to ensure the absence of arbitrage.
Nevertheless one might ask the question whether one could still work
with the wider class of $\mathcal{LSS}$ processes which are not
semimartingales.
Here we note that the standard semimartingale assumption in
mathematical finance is only valid for \textit{tradeable} assets in the
sense of assets which can be held in
a portfolio.
Hence, when dealing with, for example, electricity spot prices, this assumption
is not valid since electricity is essentially non-storable. Hence, such
a spot price
cannot be part of any financial portfolio and, therefore, the requirement
of being a martingale under some equivalent measure $Q$ is not necessary.

Guasoni, R{\'a}sonyi and
Schachermayer~\cite{GRS2008} have pointed out that, while in
frictionless markets
martingale measures play a key role, this is not the case any more in
the presence of market imperfections. In fact, in markets with
transaction costs, \emph{consistent price systems} as introduced in
Schachermayer~\cite{Schachermayer2004} are essential.
In such a set-up, even processes which are not semimartingales can
ensure that we have
\emph{no free lunch with vanishing risk} in the sense of Delbaen and
Schachermayer~\cite{DelbaenSchachermayer1994}.
It turns out that if a continuous
price process has \emph{conditional full support}, then it admits consistent
price systems for arbitrarily small transaction costs, see Guasoni, R{\'
a}sonyi and
Schachermayer~\cite{GRS2008}.
It has recently been shown by
Pakkanen~\cite{Pakkanen2010}, that under certain conditions, a
$\mathcal{BSS}$
process has conditional full support. This means that such processes
can be used in financial applications without necessarily giving rise
to arbitrage opportunities.

\subsection{Model extensions}
Let us briefly point out some model extensions concerning a
multi-factor structure, non-stationary effects, multivariate models and
alternative methods for incorporating stochastic volatility.

A straightforward extension of our model is to study a superposition
of $\mathcal{LSS}$ processes for the spot price dynamics. That is, we
could replace the process $Y$ by
a superposition of $J \in\mathbb{N}$ factors:
%
\begin{equation}
\label{sup} \sum_{i=1}^J w_i
Y_t^{(i)} \qquad\mbox{where } w_1,\ldots,
w_J \geq0, \sum_{i=1}^Jw_i
=1,
\end{equation}
and where all $Y_t^{(i)}$ are defined as in (\ref{DefYSpot}) for
independent L\'{e}vy
processes $L^{(i)}$ and independent stochastic volatility processes
$\omega^{(i)}$, in both the geometric and the arithmetic model.
Such models include
the Benth, Kallsen and
Meyer-Brandis~\cite{BenthKallsenMeyerBrandis2007} model as a special
case. A
superposition of factors $Y^{(i)}$ opens up for separate modelling of
spikes and other effects. For instance, one could let the first factor
account for the spikes, using a L\'evy process with big jumps at low
frequency, while the function $g$ forces the jumps back at a high
speed. The next factor(s) could model the ``normal'' variations of the market, where one observes a slower force of
mean-reversion, and high frequent Brownian-like noise, see Veraart and
Veraart~\cite{VV2}
for extensions along these lines.
Note that all the results we derive in this paper based on the one
factor model can be easily generalised to accommodate for the
multi-factor framework.\vadjust{\goodbreak} It should be noted that this type of
``superposition'' is quite different from the concept behind supOU
processes as studied in, for example, Barndorff-Nielsen and
Stelzer~\cite{BNStelzer2010b}.

In order to study various energy spot prices simultaneously, one can consider
extensions to a multivariate framework along the lines of
Barndorff-Nielsen and
Stelzer~\cite{BNStelzer2010,BNStelzer2010b},
Veraart and Veraart~\cite{VV2}.

In addition, another interesting aspect which we leave for future
research is the question of alternative ways of introducing stochastic
volatility in $\mathcal{VMLV}$ processes.
So far, we have introduced stochastic volatility by considering a
stochastic proportional of the
driving L\'{e}vy process, that is, we work with a stochastic integral of
$\omega$ with respect to $L$.
An alternative model specification could be based on a stochastic time
change $\int_{-\infty}^t G(t, s) \,\mathrm{d}L_{\omega_s^{2+}}$, where $\omega_s^{2+} = \int_0^s\omega_u^2\,\mathrm{d}u$.
Such models can be constructed in a fashion similar to that of
volatility modulated non-Gaussian Ornstein--Uhlenbeck processes
introduced in Barndorff-Nielsen and
Veraart~\cite{BNVeraart2011}.
We know that outside the Brownian or stable L\'{e}vy framework,
stochastic proportional and stochastic time change are not equivalent.
Whereas in the first case the jump size is modulated by a volatility
term, in the latter case the speed of the process is changed randomly.
These two concepts are in fact fundamentally different (except for the
special cases pointed out above) and, hence, it will be worth
investigating whether a combination of stochastic proportional and
stochastic time change might be useful in certain applications.

\section{Pricing of forward contracts}\label{SectionForward}

In this subsection, we are concerned with the calculation of the
forward price $F_t(T)$ at time $t \geq0$ for
contracts maturing at time $T\geq t$. We denote by $T^*<\infty$ a
finite time horizon for the forward market, meaning
that all contracts of interest mature \textit{before} this date. Note that
in energy markets, the corresponding commodity typically gets delivered
over a delivery period rather than at a fixed point in time. Extensions
to such a framework can be dealt with using standard methods, see,
for example, Benth, {\v{S}}altyt{\.e} Benth and
Koekebakker~\cite{BSBK-book} for more details.

Let $S=(S)_{t \geq0}$ denote the spot price, being either of
geometric or arithmetic kind as defined in (\ref{eqgambit-spot}) and
(\ref{eqaambit-spot}), respectively, with
\[
\overline{Y}_t=\int_{-\infty}^tG(t,s)
\omega_{s-} \,\mathrm{d}L_s , \qquad Z_t =
\omega_t^2=\int_{-\infty}^ti(t,s)
\,\mathrm{d}U_s,
\]
where the stochastic volatility $\omega$ is chosen as previously
defined in (\ref{eqambitsv}).
Clearly, the corresponding results for $\mathcal{LSS}$ processes can be
obtained by choosing
\mbox{$G(t,s)=g(t-s)$}.
We use the conventional definition of a forward price in incomplete
markets, see Duffie~\cite{duffie},
ensuring the martingale property of $t\mapsto F_t(T)$,
%
\begin{equation}
F_t(T)=\E_{Q} [S_T | \mathcal{F}_t
],\qquad  0 \leq t \leq T \leq T^*,
\end{equation}
with $Q$ being an equivalent probability measure to $P$. Here, we
suppose that $S_T\in L^1(Q)$, the space
of integrable random variables. In a moment, we shall introduce
sufficient conditions for this.\vadjust{\goodbreak}

\subsection{Change of measure by generalised Esscher transform}
In finance, one usually uses equivalent martingale measures $Q$,
meaning that the equivalent probability measure
$Q$ should turn the discounted price dynamics of the underlying asset
into a
(local) $Q$-martingale. However, as we have already discussed, this
restriction is not relevant in, for example, electricity markets since the spot
is not tradeable. Thus, we may choose any equivalent probability $Q$ as
pricing measure. In
practice, however, one restricts to a parametric class of equivalent
probability measures, and
the standard choice seems to be given by the Esscher transform, see
Benth, {\v{S}}altyt{\.e} Benth and
Koekebakker~\cite{BSBK-book}, Shiryaev~\cite{Sh}.
The Esscher transform naturally extends the Girsanov transform to L\'
evy processes.

To this end, consider $Q^{\theta}_L$ defined as the (generalised)
Esscher transform of $L$ for a
parameter $\theta(t)$ being a Borel measurable function. Following
Shiryaev~\cite{Sh} (or Benth, {\v{S}}altyt{\.e} Benth and
Koekebakker~\cite{BSBK-book}, Barndorff-Nielsen and
Shiryaev~\cite{BNShiryaev2010}), $Q^{\theta}_L$
is defined via the Radon--Nikodym density process
%
\begin{eqnarray}
\frac{\mathrm{d}Q^{\theta}_L}{\mathrm{d}P} \Big\vert_{\mathcal
{F}_t}=\exp \biggl(\int_{0}^t
\theta(s) \,\mathrm{d} L_s -\int_{0}^t
\phi_L\bigl(\theta(s)\bigr) \,\mathrm{d}s \biggr)
\end{eqnarray}
for $\theta(\cdot)$ being a real-valued function which is integrable
with respect to the L\'evy process on $[0, T^*]$,
and
\[
\phi_L(x)=\log\bigl(\mathbb{E}\bigl(\exp(xL_1)\bigr)
\bigr)=\psi(-\mathrm{i} x)= d x + \frac{1}{2} x^2 b + \int
_{\R}\bigl(\mathrm{e}^{xz}-1-xz\mathbb{I}_{\{|z|\leq1}\}
\bigr)\ell_L(\mathrm{d}z),
\]
(for $x \in\mathbb{R}$) being the log-moment generating function of
$L_1$, assuming that the moment generating function of $L_1$ exists.

A special choice is the `constant'  measure change, that is,
letting
%
\begin{equation}
\label{constantMC}\theta(t) = \theta\mathbf {1}_{[0,\infty)}(t).
\end{equation}
In this case,
if under the measure $P$, $L$ has characteristic triplet $(d,b,\ell_L)$, where $d$ is the drift, $b$ is the squared volatility of the
continuous martingale part and $\ell_L$ is the L\'evy measure in the
L\'
evy--Khinchine representation,
see
Shiryaev~\cite{Sh}, a fairly straightforward calculation shows that, see
Shiryaev~\cite{Sh} again,
the Esscher transform preserves the L\'evy property of $L$, and the
characteristic triplet under the measure $Q_L^{\theta}$ on the interval
$[0, T^*]$ becomes
$(d_{\theta},b,\exp(\theta\cdot) \ell_L)$, where
\[
d_{\theta}=d+b\theta+\int_{|z|\leq1}z\bigl({
\mathrm{e}}^{\theta z}-1\bigr) \ell_L(\mathrm{d}z) .
\]
This comes from the simple fact that the logarithmic moment generating
function of $L$ under $Q^{\theta}_L$
is
%
\begin{equation}
\label{eqmomentgen-esscher} \phi^{\theta}_L(x)\triangleq
\phi_L(x+\theta)-\phi_L(x) .
\end{equation}
%
%
\begin{rem*}
It is important to note here that the choice of $\theta(t)$ (as,
e.g., in (\ref{constantMC})) forces us to choose a \emph{starting time}
since the function $\theta$\vadjust{\goodbreak}
will \emph{not} be integrable with respect to $L$ on the unbounded
interval $(-\infty,t)$.
Recall that the only reason why we model from $-\infty$ rather than
from $0$ is the fact that we want to be able to obtain a stationary
process under the probability measure $P$.
Throughout this section, we choose the starting time to be zero, which
is a convenient choice since
$L_0=0$, and it is also practically reasonable since this can be
considered as the time from which we start to observe the process. With
such a choice, we do not introduce any risk premium for $t<0$.
\end{rem*}
In the general case, with a time-dependent parameter function $\theta
(t)$, the characteristic triplet
of~$L$ under $Q^{\theta}_L$ will become time-dependent, and hence the
L\'evy process property is lost.
Instead, $L$ will be an independent increment process (sometimes called
an additive process).
Note that if $L=B$, a Brownian motion, the Esscher transform is simply
a Girsanov change of measure where
$\mathrm{d}B_t=\theta(t) \,\mathrm{d}t+\mathrm{d}W_t$ for $0\leq t \leq T^*$ and a $Q^{\theta
}_L$-Brownian motion $W$.

Similarly, we do a (generalised) Esscher transform of $U$, the
subordinator driving the stochastic volatility model, see
(\ref{eqambitsv}). We define $Q^{\eta}_U$ to have the Radon--Nikodym
density process
\[
\frac{\mathrm{d}Q^{\eta}_U}{\mathrm{d}P} \Big\vert_{\mathcal{F}_t}=\exp \biggl(\int_{0}^t
\eta (s) \,\mathrm{d} U_s -\int_{0}^t
\phi_U\bigl(\eta(s)\bigr) \,\mathrm{d}s \biggr)
\]
for $\eta(\cdot)\in\R$ being a real-valued function which is integrable
with respect to $U$ on $[0,T^*]$,
and $\phi_U(x)=\log(\mathbb{E}(\exp(x U_1)))$ being the
log-moment generating function of $U_1$.
Since $U$ is a subordinator, we obtain
\[
\phi_U(x)=\widetilde{d} x + \int_0^{\infty}
\bigl({\mathrm {e}}^{xz}-1 \bigr) \ell_U(\mathrm{d}z),
\]
where $\widetilde d \geq0$ and $\ell_U$ denotes the L\'{e}vy measure
associated with $U$.
%
\begin{rem*}
Our discussion above on choosing a starting value applies to the
measure transform for the volatility process as well, and hence
throughout the paper we will work under the assumption that
$\theta(s) = \eta(s) = 0,$ for $s < 0$.
Note in particular, that this assumption implies that under the
risk-neutral probability measure, the characteristic triplets of $L$
and $U$ only change on the time interval $[0, T^*]$. On the interval
$(-\infty, 0)$, we have the same characteristic triplet for $L$ and $U$
as under $P$.
\end{rem*}
Choosing $\eta(t)=\eta\mathbf{1}_{[0,\infty)}(t)$, with a constant
$\eta\in\mathbb{R}$, an Esscher transform will give a characteristic triplet
$(\widetilde{d},0,\exp(\eta\cdot) \ell_U)$, which thus preserves the
subordinator property
of $(U_t)_{0\leq t\leq T^*}$ under $Q^{\eta}_U$. For the general case,
the process $U$ will be a time-inhomogeneous
subordinator (independent increment process with positive jumps). The
log-moment generating function of $U_1$ under the measure $Q^{\eta}_U$
is denoted by $\phi_U^{\eta}(x)$.

In order to ensure the existence of the (generalised) Esscher
transforms, we need some conditions.
We need that there exists a constant $c>0$ such that $\sup_{0\leq s
\leq T^*}|\theta(s)|\leq c$,
and where $\int_{|z|>1}\exp(cz)\ell_L(\mathrm{d}z)<\infty$. (Similarly, we must
have such a condition for the L\'{e}vy measure of the subordinator
driving the stochastic volatility, that is, $\ell_U$).
Also,
we must require that exponential moments of $L_1$
and $U_1$ exist. More precisely, we suppose that parameter functions
$\theta(\cdot)$ and
$\eta(\cdot)$ of the (generalised) Esscher transform are such that
%
\begin{equation}
\label{exp-int-LandU} \int_{0}^{T^*}\int
_{|z|>1}{\mathrm{e}}^{|\theta(s)| z} \ell_L(\mathrm{d}z) \,
\mathrm{d}s<\infty ,\qquad \int_{0}^{T^*}\int
_{|z|>1}{\mathrm{e}}^{|\eta(s)| z} \ell_U(\mathrm{d}z) \,
\mathrm{d}s<\infty.
\end{equation}
The exponential integrability conditions
of the L\'evy measures of $L$ and $U$ imply the existence of
exponential moments, and thus
that the Esscher transforms $Q^{\theta}_L$ and $Q^{\eta}_U$ are well defined.

We define the probability $Q^{\theta,\eta}\triangleq Q^{\theta
}_L\times
Q^{\eta}_U$ as the
class of pricing measures for deriving forward prices. In this respect,
$\theta(t)$ may be referred to as the market price of risk, whereas
$\eta(t)$ is the market price
of volatility risk. We note that a choice $\theta>0$ will put more
weight to the positive jumps in the price dynamics, and
less on the negative, increasing the ``risk'' for big upward
movements in the prices under $Q^{\theta,\eta}$.

Let us denote by $\E_{\theta,\eta}$ the expectation
operator with respect to $Q^{\theta,\eta}$, and by $\E_{\eta}$ the
expectation with respect to $Q^{\eta}_U$.

\subsubsection{Forward price in the geometric case}
Suppose that the spot price is defined by the geometric model
\[
S_t:=S_t^g=\Lambda(t)\exp(
\overline{Y}_t) ,
\]
where $\overline Y$ is defined as in (\ref{ShortNot}).
In order to have the forward price $F_t(T)$ well defined, we need to
ensure that the spot price is integrable with
respect to the chosen pricing measure~$Q^{\theta,\eta}$. We discuss
this issue in more detail in the following.

We know that $\omega$ is positive and in general not bounded since it
is defined via a subordinator. Thus,
$G(t,s)\omega_s+\theta(s)$ (for $s\leq t$) is unbounded as well.
Supposing that
$L$ has exponential moments of all orders, we can calculate as follows
using iterated expectations conditioning on
the filtration $\mathcal{G}_t$ generated by the paths of $\omega_s$,
for $s\leq t$:
\begin{eqnarray*}
\E_{\theta,\eta} [S_T ]&=&\Lambda(T)\E_{\theta,\eta
} \biggl[
\E_{\theta,\eta} \biggl[\exp \biggl(\int_{-\infty}^TG(T,s)
\omega_{s-} \,\mathrm{d}L_s \biggr) \Big| \mathcal{G}_T
\biggr] \biggr]
\\
&=&\Lambda(T)\E_{\eta} \biggl[\exp \biggl(\int_{-\infty}^0
\phi_L\bigl(G(T,s)\omega_s\bigr) \,\mathrm{d}s \biggr)
\exp \biggl(\int_{0}^T\phi^{\theta}_L
\bigl(G(T,s)\omega_s\bigr) \,\mathrm{d}s \biggr) \biggr] .
\end{eqnarray*}
To have that $S_T\in L^1(Q^{\theta,\eta})$, the two integrals must be
finite. This puts additional restrictions on the choice of
$\eta$ and the specifications of $G(t,s)$ and $i(t,s)$. We note that
when applying the Esscher transform, we must require
that $L$ has exponential moments of all orders, a rather strong
restriction on the possible class of driving L\'evy processes.
In our empirical study, however, we will later see that the empirically
relevant cases are either that $L$ is a Brownian motion or that $L$ is
a generalised hyperbolic L\'{e}vy process, which possess exponential
moments of all orders.

We are now ready to price forwards under the Esscher transform.
%
\begin{prop}
\label{propforward-generalambit}
Suppose that $S_T\in L^1(Q^{\theta,\eta})$. Then, the forward price for
$0 \leq t \leq T \leq T^*$ is given by
\[
F_t(T)=\Lambda(T)\exp \biggl(\int_{-\infty}^tG(T,s)
\omega_{s-} \,\mathrm{d}L_s \biggr) \E_{\eta}
\biggl[\exp \biggl(\int_t^T\phi_L^{\theta
}
\bigl(G(T,s)\omega_s\bigr) \,\mathrm{d}s \biggr) \Big|
\mathcal{F}_t \biggr] .
\]
\end{prop}

\subsection{Change of measure by the Girsanov transform in the
Brownian case}
As a special case, consider $L=B$, where $B$ is a two-sided standard
Brownian motion under~$P$. In this case we apply the Girsanov transform
rather than the generalised Esscher transform, and it turns
out that a rescaling of the transform parameter function $\theta(t)$ by
the volatility $\omega_t$ is
convenient for pricing of forwards. To this end, consider the Girsanov transform
%
\begin{equation}
\label{girsanovtransf} B_t=W_t+\int_0^t
\frac{\theta(s)}{\omega_{s-}} \,\mathrm{d}s \qquad \mbox{for } t \geq0,\qquad B_t =
W_t\qquad \mbox{for } t < 0,
\end{equation}
that is, we set $\theta(t) = 0$ for $t < 0$.
Supposing that the Novikov condition
\[
\E \biggl[\exp \biggl(\frac12\int_{0}^{T^*}
\frac{\theta^2(s)}{\omega^2_s} \,\mathrm{d}s \biggr) \biggr]<\infty,
\]
holds, we know that $W_t$ is a Brownian motion for $0\leq t\leq T^*$
under a probability $Q^{\theta}_B$ having
density process
\[
\frac{\mathrm{d}Q^{\theta}_B}{\mathrm{d}P} \Big\vert_{\mathcal{F}_t}=\exp \biggl(-\int_{0}^t
\frac{\theta(s)}{\omega_{s-}} \,\mathrm{d}B_s- \frac12\int_{0}^t
\frac{\theta^2(s)}{\omega^2_s} \,\mathrm{d}s \biggr) .
\]
Suppose that there exists a measurable function
$j(t)$ such that
%
\begin{equation}
\label{suff-novikov} j(t)\leq\frac{i(t,s)}{i(0,s)}
\end{equation}
for all $0\leq s\leq t\leq T^*$, with
\[
\int_{0}^{T^*}\frac{\theta^2(s)}{j(s)} \,\mathrm{d}s<
\infty.
\]
Furthermore, suppose the moment generating function of $\omega_0^{-2}$
exists on the interval $[0, C_U)$. Then,
for all $\theta(t)$ such that $0.5\int_{0}^{T^*}\theta^2(s)/j(s)
\,\mathrm{d}s\leq C_U$, the Novikov condition is satisfied, since
by the subordinator property of $U_t$ (restricting our attention to
$t\geq0$)
\[
\omega^2_t=\int_{-\infty}^ti(t,s)
\,\mathrm{d}U_s\geq\int_{-\infty}^{0}i(t,s)
\,\mathrm{d}U_s\geq j(t)\int_{-\infty}^0i(0,s)
\,\mathrm{d}U_s=j(t)w_0^2 ,
\]
and therefore
\[
\E \biggl[\exp \biggl(\frac12\int_{0}^{T^*}
\frac{\theta^2(s)}{\omega^2_s} \,\mathrm{d}s \biggr) \biggr]\leq \E \biggl[\exp \biggl(\frac12
\int_{0}^{T^*}\frac{\theta^2(s)}{j(s)} \,\mathrm{d}s
\omega_0^{-2} \biggr) \biggr]<\infty.
\]
Specifying $i(t,s)=\exp(-\lambda(t-s))$, we have that
$i(t,s)/i(0,s)=\exp(-\lambda t)=j(t)$, and
condition \eqref{suff-novikov} holds with equality.

\subsubsection{Forward price in the geometric case}
We discuss the integrability of $S_T=S_T^g$ with respect to $Q^{\theta
,\eta}\triangleq Q^{\theta}_B\times Q^{\eta}_U$.
By double conditioning with respect to the filtration generated by the
paths of $\omega_t$, we find
\begin{eqnarray*}
\E_{\theta,\eta} [S_T ]&=&\Lambda(T)\exp \biggl(\int
_{0}^{T}G(T,s)\theta(s) \,\mathrm{d}s \biggr)
\E_{\theta,\eta} \biggl[\E_{\theta,\eta
} \biggl[\exp \biggl(\int
_{-\infty}^TG(T,s)\omega_{s-} \,
\mathrm{d}W_s \biggr) \Big| \mathcal{G}_T \biggr] \biggr]
\\
&=&\Lambda(T)\exp \biggl(\int_{0}^{T}G(T,s)
\theta(s) \,\mathrm{d}s \biggr)\E_{\eta
} \biggl[\exp \biggl(\frac12\int
_{-\infty}^TG^2(T,s)\omega^2_s
\,\mathrm{d}s \biggr) \biggr] .
\end{eqnarray*}
From collecting the conditions on $G, i, \theta$ and $\eta$ for
verifying all the steps above, we find that if
$s\mapsto G(T,s)\theta(s)$ is integrable on $[0,T)$ (recall that
$\theta
(s) = \eta(s)= 0$ for $s< 0$) and $s\mapsto G^2(T,s)i(s,v)$ is
integrable on $[v,T)$ for all
$-\infty<v< T$, then $S_T\in L^1(Q^{\theta,\eta})$ as long as
%
\begin{equation}
\label{Combinedcond} \int_{-\infty}^T\int
_{|z| > 1}\exp \biggl(z \biggl\{\frac{1}{2}\int
_v^TG^2(T,s)i(s,v) \,\mathrm{d}s+\bigl|
\eta(v)\bigr| \biggr\} \biggr) \ell_U(\mathrm{d}z) \,\mathrm{d}v<\infty.
\end{equation}
We assume these conditions to hold.

We state the forward price for the case $L=B$ and the Girsanov change
of measure discussed above.
%
\begin{prop}
\label{propforward-bmambit}
Suppose that $L=B$ and that $Q^{\theta}_B$ is defined by the Girsanov
transform in \eqref{girsanovtransf}. Then,
for $0\leq t\leq T\leq T^*$,
\begin{eqnarray*}
F_t(T)&=&\Lambda(T)\exp\biggl(\int_{-\infty}^tG(T,s)
\omega_{s-} \,\mathrm{d}W_s + \frac{1}{2}\int
_{-\infty}^t\int_t^TG^2(T,s)i(s,v)
\,\mathrm{d}s \,\mathrm{d}U_v
\\
&&\hspace*{44pt}{}+\int_{0}^TG(T,s)\theta(s) \,\mathrm{d}s +\int
_t^T\phi^{\eta}_U\biggl(
\frac{1}{2}\int_v^TG^2(T,s)i(s,v)
\,\mathrm{d}s \biggr) \,\mathrm{d}v\biggr) .
\end{eqnarray*}
\end{prop}
Let us consider an example.
%
\begin{ex} In the BNS stochastic volatility model, we have $i(t,s)=\exp
(-\lambda(t-s))$.
Hence,
\[
\int_t^TG^2(T,v){
\mathrm{e}}^{-\lambda(v-s)} \,\mathrm{d}v={\mathrm {e}}^{-\lambda(t-s)}\int
_t^TG^2(T,v){\mathrm{e}}^{\lambda(t-v)}
\,\mathrm{d}v
\]
which yields,
\[
\int_{-\infty}^t\int_t^TG^2(T,v)i(v,s)
\,\mathrm{d}v \,\mathrm{d}U_s=Z_t\int
_t^TG^2(T,v){\mathrm{e}}^{\lambda(t-v)}
\,\mathrm{d}v .
\]
This implies from Proposition~\ref{propforward-bmambit} that the
forward price is affine in $Z$, the (square of the) stochastic
volatility. The stochastic volatility model studied in
Benth~\cite{B} is recovered by choosing
$G(t,s)=\exp(-\alpha(t-s))$.
\end{ex}

\subsubsection{On the case of constant volatility}
Suppose for a moment that the stochastic volatility process $\omega_t$
is identical to one (i.e., that we do not have any stochastic
volatility in the model). In this case, the forward price becomes
\begin{eqnarray*}
F_t(T)&=&\Lambda(T)\exp \biggl(\int_ {-\infty}^tG(T,s)
\,\mathrm{d}W_s+\int_{0}^TG(T,s)
\theta(s) \,\mathrm{d}s \biggr)
\\
&=&\Lambda(T)\exp \biggl(\int_{-\infty}^tG(T,s) \,
\mathrm{d}B_s+\int_t^TG(T,s)\theta
(s) \,\mathrm{d}s \biggr),
\end{eqnarray*}
where $W_t = B_t$ for $t < 0$.
Hence, the logarithmic forward (log-forward) price is
\[
\ln F_t(T)=\ln\Lambda(T)+\int_t^TG(T,s)
\theta(s) \,\mathrm{d}s+M_t(T) ,
\]
with
\[
M_t(T)=\int_{-\infty}^tG(T,s) \,
\mathrm{d}B_s
\]
for $t\leq T$. Note that $t\mapsto M_t(T)$, for $t\geq0$, is a
$P$-martingale with the property (for $S_t=S_t^g$)
\[
M_t(t)=\overline Y_t=\ln S_t-\ln\Lambda(t) .
\]
In the classical Ornstein--Uhlenbeck case, with $G(t,s)=g(t-s)$,
$g(x)=\exp(-\alpha x)$ for $\alpha> 0$, we easily compute that
\[
M_t(T)={\mathrm{e}}^{-\alpha(T-t)}\overline Y_t={\mathrm
{e}}^{-\alpha(T-t)}Y_t ,
\]
and the forward price is explicitly dependent on the current spot price.

In the general case, this does not hold true.
We have that $M_T(T)=\overline Y_T$, not unexpectedly, since the
forward price converges to the spot at maturity (at least
theoretically). However,
apart from the special time point $t=T$, the forward price will in
general \textit{not} be
a function of the current spot, but a function of the process $M_t(T)$.
Thus, at time $t$,
the forward price will depend on
\[
M_t(T)=\int_{-\infty}^tG(T,s) \,
\mathrm{d}B_s ,
\]
whereas the spot price depends on
\[
\overline Y_t=\int_{-\infty}^tG(t,s) \,
\mathrm{d}B_s .
\]
The two stochastic integrals can be pathwise interpreted (they are both
Wiener integrals
since the integrands are deterministic functions), and both $Y_t$ and
$M_t(T)$ are generated by integrating over the same paths of a Brownian
motion. However, the paths are scaled by two
different functions $G(T,s)$ and $G(t,s)$. This allows for an
additional degree of flexibility when
creating forward curves compared to affine structures.

In the classical Ornstein--Uhlenbeck case, the forward curve as a
function of time to maturity
$T-t$ will simply be a discounting of today's spot price, discounted by
the speed of mean reversion of
the spot (in addition comes deterministic scaling by the seasonality
and market price of risk). To highlight
the additional flexibility in our modelling framework of semistationary
processes, suppose for the sake of illustration that
$G(t,s)=g_1(t)g_2(s)$. Then
\[
M_t(T)=\frac{g_1(T)}{g_1(t)} \overline Y_t .
\]
If furthermore $\lim_{T\rightarrow\infty}g_1(T):=g_1(\infty)\neq
0$, we
are in a situation where the long end (i.e.,~$T$ large) of the
forward curve is not a constant. In fact, we find for $t \geq0$ that
\[
\lim_{T\rightarrow\infty} \biggl(\ln F_t(T)-g_1(t)\int
_t^Tg_2(s)\theta (s) \,\mathrm{d}s-
\ln\Lambda(T) \biggr)=\bigl(\ln S_t-\ln\Lambda(t)\bigr)
\frac
{g_1(\infty
)}{g_1(t)} .
\]
Since $\ln S_t$ is random, we will have a randomly fluctuating long end
of the forward curve.
This is very different from
the situation with a classical mean-reverting spot dynamics, which
implies a deterministic forward price in the long end (dependent on the
seasonality and market price of risk only). Various shapes of the
forward curve $T\mapsto F_t(T)$ can also be modelled via different
specifications of $G$. For instance, if $g_1(T)$ is a decreasing
function, we obtain the contango and backwardation situations depending
on the spot price being above or below the mean. If $T\mapsto g_1(T)$
has a hump, we will also observe a hump in the forward curve. For
general specifications of $G$ we can have a high degree of flexibility
in matching desirable shapes of the forward curve.

Observe that the time-dynamics of the forward price can be considered
as correlated with the spot rather than
directly depending on the spot. In the Ornstein--Uhlenbeck situation,
the log-forward price can be considered as a linear regression on the
current spot price, with time-dependent coefficients. This is not the
case for general specifications. However,
we have that $M_t(T)$ and $\overline Y_t$ are both normally distributed
random variables (recall that we are still restricting our attention to
$L=B$), and the correlation
between the two is
\[
\operatorname{Cor}\bigl(M_t(T),\overline Y_t\bigr)=\frac{\int_{-\infty}^tG(T,s)G(t,s)
\,\mathrm{d}s}{\sqrt {\int_{-\infty}^tG^2(T,s) \,\mathrm{d}s\int_{-\infty}^tG^2(t,s) \,\mathrm{d}s}} .
\]
Obviously, for $G(t,s)=g(t-s)=\exp(-\alpha(t-s))$, the correlation is
1. In conclusion, we can obtain a weaker stochastic dependency between
the spot and forward price than in the classical mean-reversion case by
a different specification of the function $G$.

\subsubsection{Affine structure of the forward price}
In the discussion above, we saw that the choice $G(t,s)=g_1(t)g_2(s)$
yielded a forward price expressible in terms of $Y_t$.
In the next proposition, we prove that this is the only choice
of $G$ yielding an affine structure. The result is slightly
generalising the analysis of
Carverhill~\cite{Carverhill2003}.
%
\begin{prop}\label{ProfForwardAffine}
The forward price in Proposition~\ref{propforward-bmambit} is affine
in $\overline Y_t$ and $Z_t$ if
there exist functions $g_1, g_2, i_1$ and $i_2$ such that
$G(t,s)=g_1(t)g_2(s)$ and $i(t,s)=i_1(t)i_2(s)$. Conversely, if the
forward price is affine in $\overline Y_t$ and $Z_t$, and $G$ and $i$
are strictly positive and continuously differentiable in the first argument,
then there exists functions $g_1, g_2, i_1$ and $i_2$ such that
$G(t,s)=g_1(t)g_2(s)$ and $i(t,s)=i_1(t)i_2(s)$.
\end{prop}
Obviously, the choice of $G$ and $i$ coming from OU-models,
\[
G(t,s)=g(t-s)=\exp\bigl(-\alpha(t-s)\bigr) ,\qquad  i(t,s)=\exp\bigl(-\lambda(t-s)
\bigr) ,
\]
satisfy the conditions in the proposition above. In fact, appealing to
similar arguments as in the proof of Proposition \ref
{ProfForwardAffine} above, one can show that this is the \textit{only}
choice (modulo multiplication by a constant) which is stationary and
gives an affine structure in the spot and volatility for the forward
price dynamics. In particular, the specification $g(x)=\sigma/(x+b)$
considered in
Example~\ref{ex-bjerksund-g} gives a stationary spot price dynamics,
but not an affine structure in the spot for the forward price.

\subsubsection{Risk-neutral dynamics of the forward price and the
Samuelson effect}
Next, we turn our attention to the risk-neutral dynamics of the forward price.
%
\begin{prop}\label{PropRiskNeutralDynamicsF}
Assume that the assumptions of Proposition~\ref{propforward-bmambit}
hold and that $Q_U^{\eta}$ is given by the (simple) Esscher transform. Then
the risk-neutral dynamics of the forward price $F_t(T)$
is given by
\begin{eqnarray*}
\frac{\mathrm{d}F_t(T)}{F_{t-}(T)} = G(T,t) \omega_{t-}\,\mathrm{d}W_t +
\int_0^{\infty} \biggl(\exp \biggl(\frac12
H_T(t, t) z \biggr)-1 \biggr)\widetilde N_U(\mathrm{d}z,\mathrm{d}t),\qquad  0
\leq t \leq T \leq T^*,
\end{eqnarray*}
where $H_T(t,t)=\int_t^TG^2(T,s)i(s,t) \,\mathrm{d}s$. Moreover
$\widetilde N_U(\mathrm{d}z,\mathrm{d}t) = N_U(\mathrm{d}z,\mathrm{d}t)-\ell_U^{\eta}(\mathrm{d}z) \,\mathrm{d}t$ is a
$Q^{\eta
}_U$-martingale, where
$N_U$ denotes the Poisson random measure associated with $U$, and $\ell_U^{\eta}= \exp(\eta\cdot)\ell_U$ is the L\'{e}vy measure of $U$
under $Q^{\eta}_U$.
\end{prop}
We observe that the dynamics will jump according to the changes in
volatility given by the process
$U_t$. As expected, the integrand in the jump expression tends to zero
when $T-t\rightarrow0$, since the forward price must (at least
theoretically) converge to the spot when time to maturity goes to zero.\vadjust{\goodbreak}

The forward dynamics will have a stochastic volatility given by
$G(T,t)\omega_{t-}$. Hence, whenever $\lim_ {t\uparrow T}G(T,t)$
exists, and $G(T,T)=1$, we have $a.s.$,
\[
\lim_{t\uparrow T}G(T,t)\omega_{t-}=\omega_{T-}.
\]
When passing to the limit, we have implicitly supposed that we work
with the version of $\omega_{t-}$ having left-continuous paths with
right-limits. By the definition of our integral in $\overline{Y}_t$,
where the integrand is supposed predictable, this can be done.
Thus, we find that the forward volatility converges to the spot
volatility as time to maturity tends to zero, which is known as the
Samuelson effect. Contrary to the classical situation where this
convergence goes exponentially, we may have
many different shapes of the volatility term structure resulting from
our general modelling framework.

In
Bjerksund, Rasmussen and
Stensland~\cite{BRS}, a forward price dynamics for electricity
contracts is
proposed to follow
%
\begin{equation}
\frac{\mathrm{d}F_t(T)}{F_t(T)}= \biggl\{a+\frac{\sigma}{T-t+b} \biggr\} \,
\mathrm{d}W_t ,
\end{equation}
where $a, b$ and $\sigma$ are positive constants. They argue that in
electricity markets, the Samuelson effect
is stronger close to maturity than what is observed in other commodity
markets, and they suggest to capture this
by letting it increase by the rate $1/(T-t+b)$ close to maturity of the
contracts. This is in contrast to the common choice of
volatility being $\sigma\exp(-\alpha(T-t))$, resulting from using the
Schwartz model
for the spot price dynamics. There is no reference to any spot model in the
Bjerksund, Rasmussen and
Stensland~\cite{BRS} model. The constant
$a$ comes from a non-stationary behaviour, which can be incorporated in
the $\mathcal{VMLV}$ framework. However, here we focus on the
stationary case and choose $a=0$. Then we see that we can model the
spot price by the $\mathcal{BSS}$ process
\[
Y_t=\int_{-\infty}^tg(t-s) \,
\mathrm{d}B_s\qquad \mbox{with } g(x)=\frac
{\sigma}{x+b}.
\]
Thus, after doing a Girsanov transform, we recover the risk-neutral
forward dynamics of
Bjerksund, Rasmussen and
Stensland~\cite{BRS}.
It is interesting to note that with this spot price dynamics, the
forward dynamics is not affine in the spot. Hence, the
Bjerksund, Rasmussen and
Stensland~\cite{BRS} model is an example of a non-affine forward dynamics.
Whenever $\sigma\neq b$, we do not have that
$g(t,t)=1$, and thus the Bjerksund, Rasmussen and
Stensland~\cite{BRS} model does not satisfy the
Samuelson effect, either.

\subsubsection{Option pricing}
We end this section with a discussion of option pricing. Let us assume
that we have given an option with exercise time $\tau$ on a forward
with maturity at time $T\geq\tau$. The option pays $f(F_{\tau}(T))$,
and we are interested in finding the price at time $t\leq\tau$, denoted
$C(t)$. From arbitrage theory, it holds that
%
\begin{equation}
C(t)={\mathrm{e}}^{-r(\tau-t)}\E_{Q} \bigl[f\bigl(F_{\tau}(T)
\bigr) | \mathcal{F}_t \bigr] ,
\end{equation}
where $Q$ is the risk-neutral probability. Choosing $Q= Q^{\theta,
\eta
}$ as coming from the Esscher transform above, we can derive option prices
explicitly in terms of the characteristic function of $U$ by Fourier
transformation.

%
\begin{prop}\label{OptionPrice}
Let $Q= Q^{\theta, \eta}$ be the probability measure obtained from the
Esscher transform. Let $p(x)=f(\exp(x))$, and suppose that $p\in
L^1(\R
)$. By applying the definitions of Fourier transforms and their
inverses in Folland~\cite{Folland}, we have that
$
p(x)=\frac1{2\pi}\int_{\R}\widehat{p}(y){\mathrm{e}}^{\mathrm
{i}xy} \,\mathrm{d}y$,
with $\widehat{p}(y)$ is the Fourier transform of $p(x)$ defined by
$
\widehat{p}(y)=\int_{\R}p(x){\mathrm{e}}^{-\mathrm{i}xy} \,\mathrm{d}x .
$
Suppose that $\widehat p \in L^1(\mathbb{R})$. Then the option price is
given by
\begin{eqnarray*}
C_t&=&{\mathrm{e}}^{-r(\tau-t)}\\
&&{}\times\frac1{2\pi}\int
_{\R}\widehat {p}(y)\exp \biggl(\mathrm {i}y \biggl(\ln
\Theta(\tau,T)+\int_{-\infty}^tG(T,s)
\omega_{s-} \,\mathrm{d}W_s-\int_{-\infty}^t
\frac{1}{2}G^2(T,s)\omega_s^2\,
\mathrm{d}s \biggr) \biggr)
\\
&&\hspace*{38pt}{} \times\exp \biggl(\int_{-\infty}^t h(T, \tau, v,
y) \,\mathrm{d}U_v \biggr) \exp \biggl(\int_t^{\tau}
\phi_U^{\eta} \bigl(h(T, \tau, v, y)\bigr) \,\mathrm{d}v
\biggr) \,\mathrm{d}y ,
\end{eqnarray*}
where $H_T(v,v) = \int_v^TG^2(T,u)i(u,v)\,\mathrm{d}u$ and
\begin{eqnarray*}
\Theta(\tau,T)&=&\Lambda(T)\exp \biggl(\int_{0}^TG(T,s)
\theta(s) \,\mathrm{d}s+\int_{\tau}^T
\phi^{\eta}_U \biggl(\frac{1}{2}H_T(v,v)
\biggr) \,\mathrm{d}v \biggr),
\\
h(T, \tau, v, y)&=& -\frac12 y^2\int_v^{\tau}G^2(T,s)i(s,v)
\,\mathrm{d}s + \mathrm{i}y \frac12 \int_{\tau
}^TG^2(T,s)i(s,v)
\,\mathrm{d}s.
\end{eqnarray*}
\end{prop}
One can calculate option prices by applying the fast Fourier transform
as long $\phi_U^{\eta}$ is known. If $p$ is not integrable (as is the
case for a call option), one may introduce a damping function to
regularize it, see Carr and Madan~\cite{CM} for details.

\subsection{The arithmetic case}
Let us consider the arithmetic spot price model,
\[
S_t:=S_t^a=\Lambda(t)+\overline{Y}_t.
\]
We analyse the forward price for this case, and discuss the affinity.
The results and discussions are reasonably parallel to the geometric
case, and we refrain from going into details but focus on some main results.

Under a natural integrability condition of the spot price with respect
to the Esscher transform measure $Q^{\theta,\eta}$, we find the
following forward price for the arithmetic model.
%
\begin{prop}
\label{propforward-arithmetic}
Suppose that $S_T\in L^1(Q^{\theta,\eta})$. Then, the forward price is
given as
\[
F_t(T)=\Lambda(T)+ \biggl\{\int_{-\infty}^tG(T,s)
\omega_{s-} \,\mathrm{d}L_s+\mathbb {E}_{\theta}[L_1]
\int_t^TG(T,s)\mathbb{E}_{\eta}[
\omega_{s} | \mathcal {F}_t] \,\mathrm{d}s \biggr\}.
\]
\end{prop}
The price is reasonably explicit, except for the conditional
expectation of the stochastic volatility $\omega_s$. By the same
arguments as in Proposition~\ref{ProfForwardAffine}, the forward price
becomes affine in the spot (or in $\overline{Y}_t$) if and only if
$G(t,s)=g_1(t)g_2(s)$ for sufficiently regular functions $g_1$ and $g_2$.

In the case $L=B$, we can obtain an explicit forward price when using
the Girsanov transform as in (\ref{girsanovtransf}). We easily compute
that the forward price becomes
%
\begin{equation}
F_t(T)=\Lambda(T)+ \biggl\{\int_{-\infty}^tG(T,s)
\omega_{s-} \,\mathrm{d}W_s+\int_{0}^TG(T,s)
\theta(s) \,\mathrm{d}s \biggr\} .
\end{equation}
We note that there is no explicit dependence of the spot volatility
$\omega_s$ except indirectly in the stochastic integral. This is in
contrast to the L\'evy case with Esscher transform. The dynamics of the
forward price becomes
%
\begin{equation}
\mathrm{d}F_t(T)=G(T,t)\omega_{t-} \,\mathrm{d}W_t .
\end{equation}

If we furthermore let $G(t,s)=g_1(t)g_2(s)$ for some sufficiently
regular functions $g_1$ and $g_2$, we find that
%
\begin{equation}
F_t(T)=\Lambda(T)+\frac{g_1(T)}{g_1(t)} \bigl(S_t-
\Lambda(t)\bigr)+\int_{t}^TG(T,s)\theta(s) \,
\mathrm{d}s .
\end{equation}
Hence, the forward curve moves stochastically as the deseasonalised
spot price, whereas the shape of the curve is deterministically given by
$g_1(T)/g_1(t)$. This shape is scaled stochastically by the
deseasonalised spot price. In addition, there is a deterministic term
which is derived from the market price of risk $\theta$.

We finally remark that also in the arithmetic case one may derive
expressions for the prices of options that are computable by fast
Fourier techniques.

\section{Empirical study}\label{Emp}
In this section, we will show the practical relevance of our new model
class for modelling empirical energy spot prices. Here we will focus on
electricity spot prices and
we will illustrate that they can be modelled by $\mathcal{LSS}$
processes -- an important subclass of $\mathcal{VMLP}$ processes.
Note that the data analysis is exploratory in nature since the
estimation theory for $\mathcal{VMLP}$ or $\mathcal{LSS}$ processes has
not been fully established yet.

\subsection{Data description}
\begin{figure}

\includegraphics{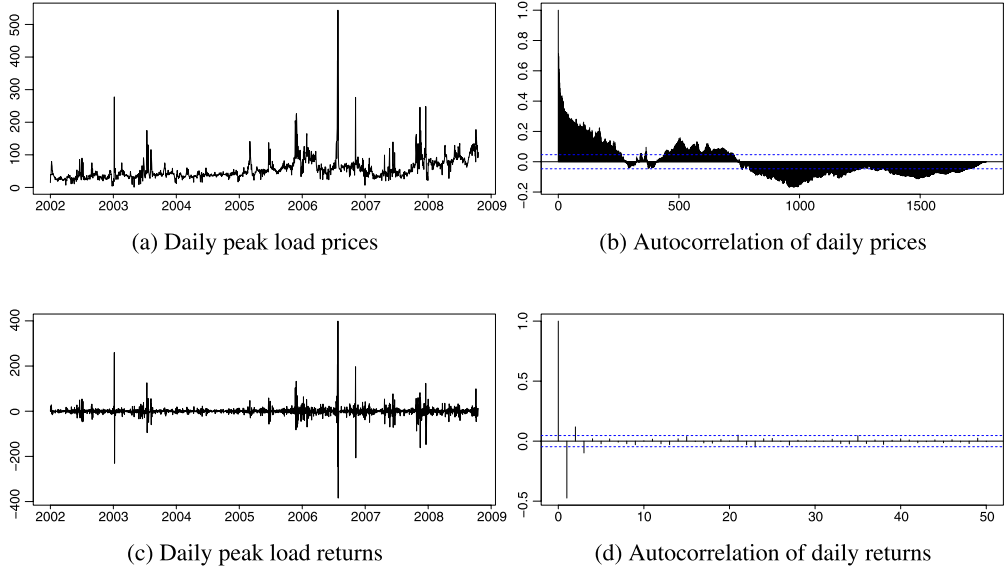}

\caption{Daily electricity peak load spot prices in Euro/MWh from the
EEX, recorded from 01.01.2002 to 21.10.2008.} \label{eldata}
\end{figure}

We study electricity spot prices
from the European Energy Exchange (EEX). We work with the daily Phelix
peak load data (i.e., the daily averages of the hourly spot prices for
electricity delivered during the 12 hours between 8am and 8pm) with
delivery days from 01.01.2002 to 21.10.2008.
Note that peak load data do not include weekends, and in total we have
1775 observations.
The daily data, their returns and the corresponding autocorrelation
functions are depicted in Figure~\ref{eldata}.

%

%

\subsection{Deseasonalising the data}
\begin{figure}

\includegraphics{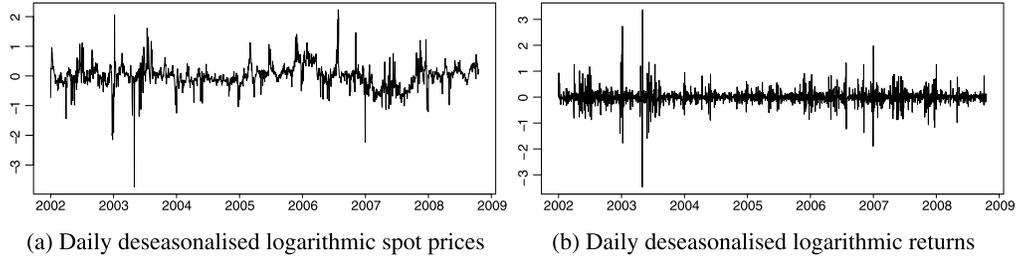}

\caption{Daily deseasonalised logarithmic spot prices.} \label{GDeseas}
\end{figure}

Before analysing the data, we have deseasonalised the spot prices.
Here, we have worked with a geometric model, that is, $S_t^{g}=\Lambda(t)
\exp(\overline Y_t)$. Then
$\log(S_t^{g})=\log(\Lambda(t))+ \overline Y_t$ where,
as suggested in, for example, Kl{\"u}ppelberg, Meyer-Brandis and
Schmidt~\cite{KMBS2010},
\begin{eqnarray*}
\log\bigl(\Lambda(t)\bigr):= \beta_0 +\beta_1 \cos
\biggl(\frac{\tau_1+2\pi t}{261} \biggr) +\beta_2 \cos \biggl(
\frac{\tau_2+2\pi t}{5} \biggr) +\beta_3 t,
\end{eqnarray*}
which takes weakly and yearly effects and a linear trend into account.
In order to ensure that the spikes do not have a big impact on
parameter estimation, we have worked with a robust estimation technique
based on
iterated reweighted least squares.
We have then subtracted the estimated seasonal function from the
logarithmic spot prices from the time series and have worked with the
deseasonalised data for the remaining part of the Section. Figure \ref
{GDeseas} depicts the deseasonalised logarithmic prices and the
corresponding returns.
%

%
\subsection{Stationary distribution of the prices}
The class of $\mathcal{VMLV}$ processes is very rich and hence in a
first step we checked whether we can restrict it to a smaller class in
our empirical work.
We have carried out unit root tests, more precisely the augmented
Dickey--Fuller test (where the null hypothesis is that a unit root is
present in the time series versus the alternative of a stationary time series);
we obtained a $p$-value which is smaller than 0.01 and, hence, clearly
reject the unit root hypothesis at a high significance level.
Also the Phillips--Perron test led to the same conclusion. Hence, in
the following, we assume that $\overline Y_t = Y_t$ is an $\mathcal
{LSS}$ process.

Next, we study the question which distribution describes the
stationary distribution of $Y$ appropriately. We know that in the
absence of stochastic volatility an $\mathcal{LSS}$ process is a moving
average process driven by a L\'{e}vy process and hence the integral is
itself infinite divisible. We are hence dealing with a stationary
infinitely divisible stochastic process, see Rajput and Rosi{\'n}ski~\cite{RajRos89},
Sato~\cite{S},
Barndorff-Nielsen~\cite{BN2011}
for more details.
The literature on spot price modelling suggest to use semi-heavy and,
in some cases, even heavy-tailed distributions in order to account for
the extreme spikes in electricity spot prices, see,
for example, Kl{\"u}ppelberg, Meyer-Brandis and
Schmidt~\cite{KMBS2010} and Benth \textit{et~al.}~\cite
{BenthKlueppelberVos} who suggested to
use the stable distribution for modelling electricity returns.

Here we focus on a mixture of a normal distribution
in the sense of mean-variance mixtures, see Barndorff-Nielsen, Kent and
S{\o}rensen~\cite{BNKentSorensen1982}.
In particular, we will focus on the
generalised hyperbolic (GH) distribution, see Barndorff-Nielsen and
Halgreen~\cite{BNHalgreen1977},
Barndorff-Nielsen~\cite{BN1977},
Barndorff-Nielsen~\cite{BN1978}, which turns out to provide a good fit
to the deseasonalised logarithmic spot prices as we will see in the following.

\subsubsection{The generalised hyperbolic distribution}
A detailed review of the generalised hyperbolic distribution
can be found in, for example, McNeil, Frey and
Embrechts~\cite{McNeilFreyEmbrechts2005} and details on the
corresponding implementation in {\tt R} based on the {\tt ghyp} package
is provided in Breymann and
L{\"u}thi~\cite{BreymannLuthi2010}.

Let $d, k \in\N$ and let ${\mathbf X}$ denote a $k$-dimensional
random vector. ${\mathbf X}$ is said to have multivariate generalised
hyperbolic (GH) distribution if
\[
{\mathbf X} \stackrel{\mathrm{law}} {=} \bolds{\mu} + W \bolds {\gamma} +
\sqrt{W} {\mathbf A} {\mathbf Z},
\]
where ${\mathbf Z}\sim N(\mathbf{0}, I_k)$, ${\mathbf A} \in\R^{d\times k}$, $\bolds{\mu}, \bolds{\gamma} \in\R^d$.
Further, $W\geq0$ is a one-dimensional random variable, independent of
$\mathbf{Z}$ and with Generalised Inverse Gaussian (GIG) distribution,
that is, $W\sim \operatorname{GIG}(\lambda, \chi, \psi)$.
The density of the GIG distribution with parameters $(\lambda, \chi,
\psi)$ is given by
\begin{eqnarray*}
f_{\mathrm{GIG}}(x) = \biggl(\frac{\psi}{\chi} \biggr)^{{\lambda}/{2}}
\frac
{x^{\lambda-1}}{2K_{\lambda}(\sqrt{\chi\psi})}\exp \biggl(-\frac
{1}{2} \biggl( \frac{\chi}{x}+\psi
x \biggr) \biggr),
\end{eqnarray*}
where $K_{\lambda}$ denotes the modified Bessel function of the third
kind, and
the parameters have to satisfy one of the following three restrictions
\begin{eqnarray*}
\chi> 0, \psi\geq0, \lambda< 0\qquad \mbox{or}\qquad \chi> 0, \psi> 0, \lambda= 0\qquad
\mbox{or}\qquad \chi\geq0, \psi> 0, \lambda> 0.
\end{eqnarray*}
Typically, we refer to $\bolds{\mu}$ as the location parameter, to
$\bolds{\Sigma}=\mathbf{A}\mathbf{A}'$ as the dispersion matrix
and to $\bolds{\gamma}$ as the symmetry parameter (sometimes also
called skewness parameter).
The parameters $\lambda, \chi, \psi$ of the GIG distribution determine
the shape of the GH distribution.
The parametrisation described above is the so-called
$(\lambda,\chi, \psi, \mu, \bolds{\Sigma}, \gamma
)$-parametrisation of the GH distribution. However, for estimation
purposes this parametrisation causes an identifiability problem and
hence we worked with the
so-called
$(\lambda, \overline\alpha, \mu, \bolds{\Sigma}, \gamma
)$-parametrisation
in our empirical study.
Note that the $(\lambda,\chi, \psi, \mu, \bolds{\Sigma}, \gamma
)$-parametrisation can be obtained by
from $(\lambda, \overline\alpha, \mu, \bolds{\Sigma}, \gamma
)$-parametrisation by setting
\[
\psi= \overline\alpha\frac{K_{\lambda+1}(\overline\alpha
)}{K_{\lambda}(\overline\alpha)},\qquad \chi= \frac{\overline\alpha^2}{\psi} = \overline
\alpha\frac{K_{\lambda}(\overline\alpha
)}{K_{\lambda+1}(\overline\alpha)},
\]
and $\lambda, \bolds{\Sigma}, \gamma$ remain the same,
see Breymann and
L{\"u}thi~\cite{BreymannLuthi2010} for more details.

\subsubsection{Estimation results}
\begin{figure}[b]

\includegraphics{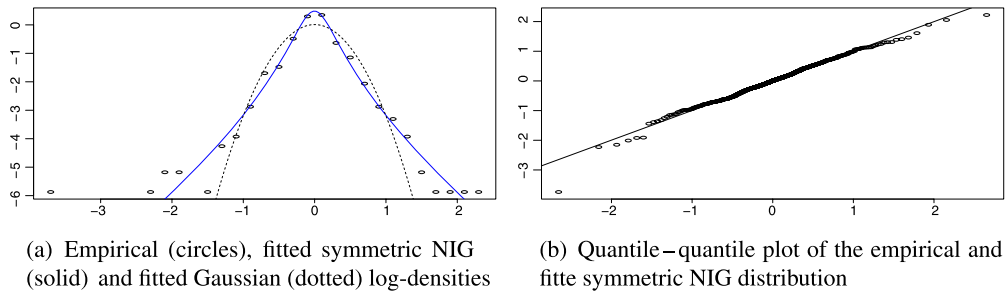}

\caption{Diagnostic plots for the estimated symmetric normal inverse
Gaussian distribution.}\label{GSymNIG}
\end{figure}

In our empirical study, we work with the one-dimensional GH
distribution. That is, $d=k=1$ and $\mu, \gamma$ and $\bolds{\Sigma}
=\sigma$ are scalars rather than a matrix and vectors, respectively.
We have fitted 11 distributions within the GH class to the
deseasonalised log-spot prices using quasi-maximum likelihood estimation:
The asymmetric and symmetric versions of the
\begin{itemize}
\item generalised hyperbolic distribution (GHYP): $\lambda\in\R,
\overline\alpha> 0$, ($\lambda\in\R, \chi> 0, \psi> 0$),
\item normal inverse Gaussian (NIG) distribution: $\lambda=-\frac
{1}{2}, \overline\alpha> 0$, ($\lambda=-\frac{1}{2}, \chi> 0, \psi
> 0$),
\item Student-$t$ distribution (with $\nu$ degrees of freedom):
$\lambda=-\nu/2 < -1, \overline\alpha= 0$, ($\lambda< 0, \chi> 0,
\psi= 0$),
\item hyperbolic distribution (HYP): $\lambda=(d+1)/2, \overline
\alpha> 0$, ($\lambda=(d+1)/2, \chi> 0, \psi> 0$),
\item Variance gamma distribution (VG): $\lambda> 0, \overline\alpha
= 0$, ($\lambda> 0, \chi= 0, \psi> 0$),
\end{itemize}
and the Gaussian distribution.
We have compared these distributions using the Akaike information
criterion, see Table~\ref{AICFit}, which suggests that the symmetric
NIG distribution is the preferred choice for the stationary
distribution of the deseasonalised logarithmic spot prices.
The diagnostic plots of the empirical and fitted logarithmic densities
and the quantile--quantile plots of the fitted symmetric NIG
distribution are depicted in Figure~\ref{GSymNIG}. We see that the fit
is reasonable.
%
\begin{table}
\caption{Model selection based on the Akaike information criterion
within the class of generalised hyperbolic distributions. We compare
both the asymmetric and the symmetric versions of the generalised
hyperbolic (GHYP), normal inverse Gaussian (NIG), Student-t, hyperbolic
(HYP), variance Gamma (VG) and the Gaussian distribution}\label{AICFit}
\begin{tabular*}{\textwidth}{@{\extracolsep{\fill}}lld{2.3}d{1.3}d{2.3}cd{2.3}cd{4.2}@{}}
\hline
{Model} & {Symmetric} & \multicolumn{1}{l}{${\widehat\lambda}$}&\multicolumn{1}{l}{${\widehat{\overline\alpha}}$}&
\multicolumn{1}{l}{${\widehat\mu}$}&\multicolumn{1}{l}{${\widehat\sigma}$}& \multicolumn{1}{l}{${\widehat\gamma}$} &
\multicolumn{1}{l}{{AIC}}&
\multicolumn{1}{l@{}}{{Log-Likel.}}\\
\hline
NIG&TRUE&-0.5&0.431&-0.001&0.395&0&1313.14&-653.57\\
GHYP&TRUE&-0.183&0.438&-0.001&0.392&0&1314.13&-653.06\\
NIG&FALSE&-0.5&0.431&-0.003&0.395&0.002&1315.10&-653.55\\
GHYP&FALSE&-0.184&0.438&-0.002&0.392&0.002&1316.10&-653.05\\
Student-t&TRUE&-1.366&0&-0.001&0.458&0&1327.28&-660.64\\
Student-t&FALSE&-1.365&0&-0.002&0.458&0.002&1329.26&-660.63\\
HYP&TRUE&1&0.150&0.000&0.375&0&1331.38&-662.69\\
HYP&FALSE&1&0.147&0.003&0.375&-0.003&1333.33&-662.66\\
VG&TRUE&0.975&0&0.003&0.379&0&1333.85&-663.92\\
VG&FALSE&0.970&0&0.007&0.379&-0.007&1335.42&-663.71\\
Gaussian&TRUE&\multicolumn{1}{c}{NA}&\multicolumn{1}{c}{Inf}&-0.000&0.395&0&1742.94&-869.47\\
\hline
\end{tabular*}
\end{table}

%

%
%
%

\subsection{Stationary $\mathcal{BSS}$ processes with generalised
hyperbolic marginals}

In our empirical study, we have seen that the symmetric normal inverse
Gaussian distribution fits the marginal distribution of the
deseasonalised logarithmic electricity prices well.
Hence, it is natural to ask whether there is
a stationary $\mathcal{BSS}$ or $\mathcal{LSS}$ process with marginal
normal inverse Gaussian or, more generally,
generalised hyperbolic distribution? The answer is yes, as we will show
in the following.
Note that the following investigation extends the study of
Barndorff-Nielsen and
Shephard~\cite{BNS2001a}, where the background driving process of an
Ornstein--Uhlenbeck process was specified, given a marginal infinitely
divisible distribution.

Let us focus on a particular $\mathcal{BSS}$ process given by
%
\begin{equation}
\label{SpecialBSS} Y_t = \mu+ c\int_{-\infty}^tg(t-s)
\omega_s\,\mathrm{d}B_s+ \gamma\int_{-\infty
}^t
q(t-s) \omega_s^2 \,\mathrm{d}s
\end{equation}
for constants $c, \gamma\in\R$ and
for stationary $\omega$ and a standard Brownian motion $B$ independent
of $\omega$.
%
\begin{rem*}
Note that we have introduced a drift term in the $\mathcal{BSS}$
process again in order to derive the general theoretical result. For
our empirical example, however, it would be sufficient to set $\gamma
=0$ as suggested by our estimation results above.
\end{rem*}
The conditional law of $Y_{t}$ given $\omega$ is normal:%
\[
Y_{t}|\omega\stackrel{\mathrm{law}} {=}N \biggl( \mu+\gamma\int
_{-\infty
}^{t}q ( t-s ) \omega_{s}^{2}
\,\mathrm{d}s,c^2\int_{-\infty
}^{t}g^{2}
( t-s ) \omega_{s}^{2}\,\mathrm{d}s \biggr) .
\]
Now suppose that $\omega^{2}$ follows an $\mathcal{LSS}$ process given
by%
\[
\omega_{t}^{2}=\int_{-\infty}^{t}i^*
( t-s )\,\mathrm{d}U_{s},
\]
where $U$ is a subordinator. Then, by a stochastic Fubini theorem we
find%
\begin{eqnarray*}
\int_{-\infty}^{t}q ( t-s ) \omega_{s}^{2}
\,\mathrm{d}s =\int_{-\infty}^{t}\int
_{u}^{t}q ( t-s ) i^* ( s-u )\,\mathrm{d}s
\,\mathrm{d}U_{u} =\int_{-\infty}^{t}k ( t-u )\,
\mathrm{d}U_{u},
\end{eqnarray*}
where $k=q\ast i^*$, the convolution of $q$ and $i^*$. Similarly,
\[
\int_{-\infty}^{t}g^{2} ( t-s )
\omega_{s}^{2}\,\mathrm{d}%
s=\int
_{-\infty}^{t}m ( t-u )\,\mathrm{d}U_{u},
\]
with $m=g^{2}\ast i^*$.
Let $\ga ( t;\newnu,\newlambda ) $ denote the gamma
density with
parameters $\newnu>0$ and $\newlambda>0$, that is,%
\[
\ga ( t;\newnu,\newlambda ) =\frac{\newlambda^{\newnu
}}{\Gamma (
\newnu ) } t^{\newnu-1}\mathrm{e}^{-\newlambda t}.
\]
Now we define
%
\begin{eqnarray}
\label{gfct} g(t)= \biggl(\newlambda\frac{\Gamma(2\newnu-1)}{\Gamma(\newnu
)^2} \biggr)^{-1/2}2^{\newnu}
\ga\biggl(t; \newnu, \frac{\newlambda}{2}\biggr) = \frac{\newlambda^{\newnu-1/2}}{\Gamma(2\newnu-1)^{1/2}}
t^{\newnu
-1}\exp \biggl(-\frac{\newlambda}{2} t \biggr)
\end{eqnarray}
for $\newnu> \frac{1}{2}$, which ensures the existence of the integral
\eqref{SpecialBSS};
then we have
\[
g^2(t)=\frac{\newlambda^{2\newnu-1}}{\Gamma(2\newnu-1)} t^{(2\newnu
-1)-1}\exp (-\newlambda t )=
\ga(t;2\newnu-1,\newlambda).
\]
Hence, if, for $\frac{1}{2}< \newnu<1$,
\[
i^*(t)= \frac{1}{\newlambda} \ga(t; 2-2\newnu,\newlambda),
\]
and if, moreover,%
\[
q ( t ) =\ga ( t;2\newnu-1,\newlambda ),
\]
we obtain%
\[
k ( t ) =m(t)=\mathrm{e}^{-\newlambda t}.
\]
In other words,%
\[
Y_{t}|\omega\stackrel{\mathrm{law}} {=}N \bigl( \mu+\gamma
\sigma_{t}^{2},c^2\sigma_{t}^{2}
\bigr),
\]
where%
\[
\sigma_{t}^{2}=\int_{-\infty}^{t}\mathrm{e}^{-\newlambda ( t-u )
}\,
\mathrm{d}%
U_{u}.
\]
We define the subordinator $\overline U$ with L\'{e}vy measure $\ell_{\overline U}$ by
$\overline U_t = U_{t/\newlambda}$. Then
\[
\sigma_{t}^{2}=\int_{-\infty}^{t}\mathrm{e}^{-\newlambda ( t-u )
}\,
\mathrm{d}%
\overline U_{\newlambda u}.
\]
Then one can easily show that the marginal distribution of $\sigma^2$
does not depend on $\newlambda$, and the parameter $\newlambda$
determines the autocorrelation structure of $\sigma^2$.

It follows that if the subordinator $\overline U$ is such that $\sigma_{t}^{2}$ has
the generalised inverse Gaussian law $\operatorname{GIG}(\lambda, \chi, \psi)$ then
the law of $%
Y_{t}$ is the generalised hyperbolic $\operatorname{GH}(\lambda, \chi, \psi, \mu,
c^2, \gamma)$.

Is there such a subordinator? The answer is yes. To see this, let
$\theta\geq0$ and note that $%
\sigma_{t}^{2}$ is infinitely divisible with kumulant function
\begin{eqnarray*}
\bar{K}\bigl\{\theta\ddagger\sigma_{t}^{2}\bigr\}&=&\log \bigl(
\mathbb {E} \bigl(\exp\bigl(-\theta\sigma_{t}^{2} \bigr)\bigr)
\bigr) 
=\log \biggl(\mathbb{E} \biggl(\exp\biggl(-
\theta\int_{-\infty}^t \mathrm{e}^{-\newlambda
(t-u)}\,\mathrm{d}
\overline U_{\newlambda u} \biggr) \biggr)\biggr)
\\
& =&\int_{0}^{\infty}\bar{K}
\bigl\{\theta \mathrm{e}^{-\newlambda u}\ddagger\overline U_{1}\bigr\}\newlambda
\,\mathrm{d}u =\int_{0}^{\infty}\bar{K}\bigl\{\theta
\mathrm{e}^{- u}\ddagger\overline U_{1}\bigr\}\,\mathrm{d}u.
\end{eqnarray*}
On the other hand, the subordinator $\overline U$ (here assumed to have
no drift) has kumulant function
\[
\bar{K}\{\theta\ddagger\overline U_{1}\} =\log \bigl(\mathbb{E} \bigl(
\exp(-\theta\overline U_{1} ) \bigr)\bigr) =-\int_{0}^{\infty}
\bigl(1-\mathrm{e}^{-\theta x}\bigr)\ell_{\overline U} ( \mathrm{d}x ),
\]
where $\ell_{\overline U}$ is the L\'{e}vy measure of $\overline U$.
Combining we find%
\[
\bar{K}\bigl\{\theta\ddagger\sigma_{t}^{2}\bigr\} =-\int
_{0}^{\infty} \bigl( 1-\mathrm{e}^{-\theta y} \bigr) \int
_{0}^{\infty}\ell_{\overline U}\bigl(\mathrm{e}^{
u}\,
\mathrm{d}y\bigr)\,\mathrm{d%
}u.
\]
That is, the L\'{e}vy measure $\ell_{\sigma^2}$ of $\sigma_{t}^{2}$ is
%
\begin{equation}
\ell_{\sigma^2}(\mathrm{d}y) = \int_{0}^{\infty}
\ell_{\overline U}\bigl(\mathrm{e}^{u}\,\mathrm{d}y\bigr)\,\mathrm{d}u. \label{wv}
\end{equation}
Thus, the question is: Does there exist a L\'{e}vy measure $\ell_{\overline U}$ on $%
\mathbb{R}_{+}$ such that $\ell_{\sigma^2}$ given by (\ref{wv}) is the
L\'{e}vy measure of
the $\operatorname{GIG}(\lambda,\chi,\psi)$ law. That, in fact, is the case since
the $\operatorname{GIG}$
laws are self-decomposable, cf. Halgreen~\cite{Halgreen1979} and Jurek
and Vervaat~\cite{JurekVervaat1983}.

\subsubsection{Implied autocorrelation structure}
Next, we focus on the autocorrelation structure implied by the choice
of the kernel functions which lead to a marginal GH distribution of the
$\mathcal{BSS}$ process.

%
\begin{prop}\label{PropCorGammaKernel}
Let $Y$ be the $\mathcal{BSS}$ process defined in the previous
subsection with kernel function $g$ as defined in \eqref{gfct}.
In the case when $\gamma=0$ and $\newnu>\frac{1}{2}$, we have
\begin{eqnarray*}
\operatorname{Cor}(Y_t,Y_{t+h})=\frac{1}{2^{\newnu-{3}/{2}} \Gamma
(\newnu
-{1}/{2} )} \bar K_{\newnu-{1}/{2}}
\biggl(\frac{\newlambda h}{2} \biggr)\qquad\mbox{for } h > 0,
\end{eqnarray*}
where $\bar{K}_{\newnu} ( x ) =x^{\newnu}K_{\newnu} (
x ) $ and $K_{\newnu}$ denotes the modified Bessel function of
the third kind.
\end{prop}

\begin{figure}

\includegraphics{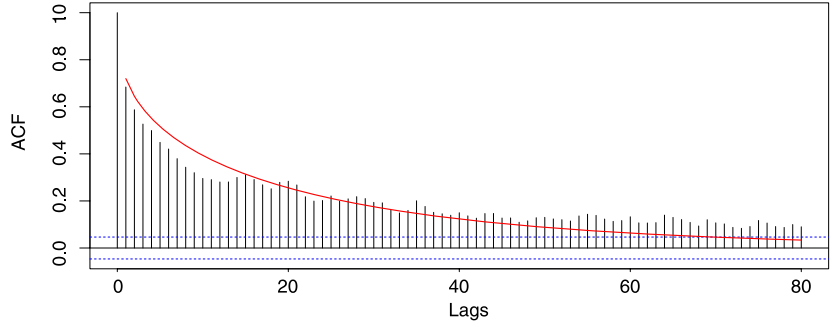}

\caption{Empirical and estimated autocorrelation function using the
gamma kernel function with $\widehat\newlambda=0.055$ and $\widehat
\newnu= 0.672$.}\label{GammaAcf}
\end{figure}

We have estimated the parameters $\newnu$ and $\newlambda$ using a
linear least squares estimate based on the empirical and the
theoretical autocorrelation function using the first $\lfloor\sqrt {1775}\rfloor=42$ lags. We obtain $\widehat\newlambda=0.055$ and
$\widehat\newnu= 0.672$. Figure~\ref{GammaAcf} shows the empirical and
the corresponding fitted autocorrelation function.

%

%
%
\begin{rem*}
Note that the estimate $\widehat\newnu= 0.672$ implies that the
corresponding $\mathcal{BSS}$ process is not a semimartingale, see,
for example, Barndorff-Nielsen and
Schmiegel~\cite{BNSch09} for details. In the context of electricity prices,
this does not need to be a concern since the electricity spot price is
not tradeable.
\end{rem*}
We observe that the autocorrelation function induced by the
gamma-kernel mimics the behaviour of the empirical autocorrelation
function adequately. However, it
does not fit the first 10 lags as well as, for example, the CARMA-kernel which
we have fitted in the following subsection, but performs noticeably
better for higher lags. The fit could be further improved by choosing
$\sigma_t^2$ to be a GIG supOU process rather than a GIG OU process.
Then one obtains an even more flexible autocorrelation structure.

\subsection{Empirical performance of a CARMA model}
\begin{figure}

\includegraphics{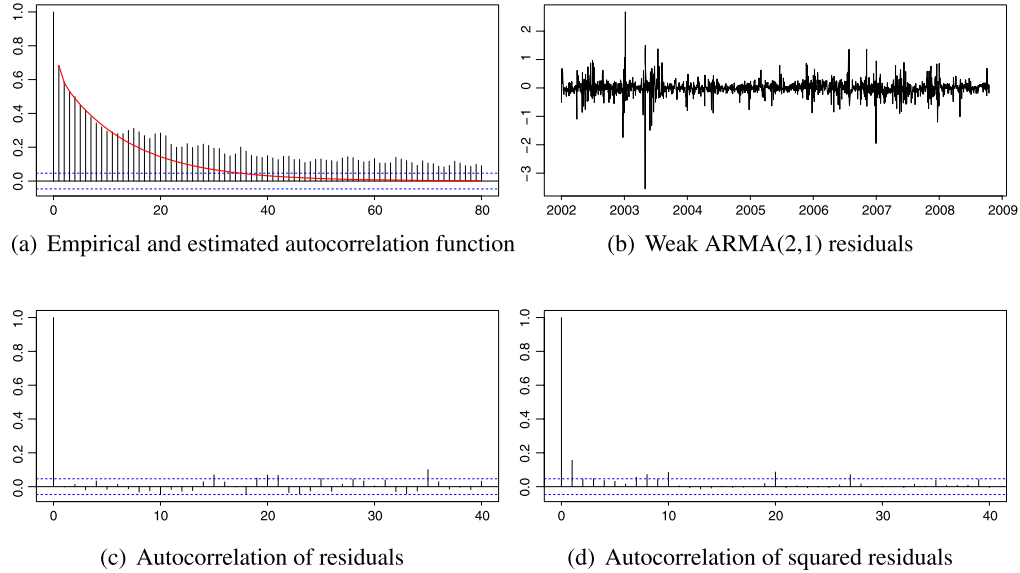}

\caption{Diagnostic plots for the estimated $\operatorname{CARMA}(2,1)$ model.
}\label{GCARMA}
\end{figure}

The recent literature on modelling electricity spot prices has
advocated the use of linear models, that is, CARMA models, as described in
detail in Section~\ref{SectUnifying}. Since CARMA models are special
cases of our general modelling framework, we briefly demonstrate their
empirical performance as well.
It is well known, see, for example, Brockwell, Davis and
Yang~\cite{BrockwellDavisYang2011}, that a
discretely sampled $\operatorname{CARMA}(p,q)$ process (for $p>q$) has a weak
ARMA($p$, $p-1$) representation. An automatic model selection using the
Akaike information criterion within the class of (discrete-time ARIMA)
models suggests that an ARMA(2,1) model is the best choice for our
data. We take that result as an indication that a $\operatorname{CARMA}(2,1)$ process
(which has a weak ARMA(2,1) representation) might be a good choice.
However, it should be noted that the relation between model selection
in discrete and continuous time still needs to be explored in detail.
We have estimated the parameters of the kernel function $g$ which
corresponds to a $\operatorname{CARMA}(2,1)$ process using quasi-maximum-likelihood
estimation based on the weak ARMA(2,1) representation.
Diagnostic plots for the estimated $\operatorname{CARMA}(2,1)$ model are provided in Figure
\ref{GCARMA}. First, we compare the empirical and the estimated
autocorrelation function, see Figure
\ref{GCARMA}(a).
Recall that the autocorrelation of $Y$ is given by (\ref{Cor}) and it
simplifies to
\[
\operatorname{Cor}(Y_{t+h}, Y_t)=\frac{ \int_0^{\infty} g(x+h)g(x) \,\mathrm{d}x }{ \int_0^{\infty} g(x)^2 \,\mathrm{d}x },
\]
if either the driving L\'{e}vy process has zero mean or if the
stochastic volatility process has zero autocorrelation.
After deseasonalising (which also includes detrending) the data, we
have obtained data which have approximately zero mean.
%
%
%
The empirical and the estimated autocorrelation function implied by a
$\operatorname{CARMA}(2,1)$ kernel function $g$ match very well for the first 12 lags.
Higher lags were however slightly better fitted by the gamma kernel
used in the previous subsection.
Figure
\ref{GCARMA}(b) depicts the corresponding residuals from the weak
ARMA(2,1) representation and Figures
\ref{GCARMA}(c)
and
\ref{GCARMA}(d) show the autocorrelation functions of the corresponding
residuals and squared residuals. Overall, we see that the fit provided
by the $\operatorname{CARMA}(2,1)$ kernel function is acceptable.

Note that in addition to estimating the parameters of the $g$ function
coming from a CARMA process one can also
recover the driving L\'evy process of a CARMA process based on recent
findings by Brockwell, Davis and
Yang~\cite{BrockwellDavisYang2011}.
This will make it possible to also address the question of whether
stochastic volatility is needed to model electricity spot prices or not.
See Veraart and Veraart~\cite{VV2} for
empirical work along those lines in the context of electricity spot
prices, whose results suggest that stochastic volatility is indeed
important for modelling electricity spot prices.

\section{Conclusion}\label{SectionConclusion}
This paper has focused on \emph{volatility modulated L\'{e}vy-driven
Volterra} ($\mathcal{VMLV}$) processes
as the building block for modelling energy spot prices.
In particular, we have introduced the class of \emph{L\'{e}vy
semistationary} ($\mathcal{LSS}$) processes as an important subclass of
$\mathcal{VMLV}$ processes, which reflect the stylised facts of
empirical energy spot prices well. This modelling framework is built on
four principles.
First, deseasonalised spot prices can be modelled directly in
stationarity to reflect the empirical fact that spot prices are
equilibrium prices determined by supply and demand and, hence, tend to
mean-revert (in a weak sense) to a long-term mean.
Second, stochastic volatility is regarded as a key factor for
modelling (energy) spot prices.
Third, our new modelling framework allows for the possibility of jumps
and extreme spikes.
Fourth, we have seen that $\mathcal{VMLP}$ and, in particular,
$\mathcal
{LSS}$ processes feature great flexibility in terms of modelling the
autocorrelation function and the Samuelson effect.

We have demonstrated that $\mathcal{VMLV}$ processes are highly
analytically tractable; we have
derived explicit formulae for the energy forward prices based on our
new spot price models, and we have shown how the kernel function
determines the Samuelson effect in our model. In addition,
we have discussed option pricing based on transform-based methods.

An exploratory data analysis on electricity spot prices shows the
potential our new approach has and more detailed empirical work is left
for future research.
Also,
we plan to address the question of model estimation and inference. It
will be important to study efficient estimation schemes for fully
parametric specifications of $\mathcal{VMLV}$- and, in particular,
$\mathcal{LSS}$-based models.

%
\begin{appendix}
\section*{Appendix: Proofs}\label{SectionProofs}
\begin{pf*}{Proof of Proposition~\ref{ThmSM}}
In order to prove the semimartingale conditions suppose for the moment
that $Y$ is a semimartingale, so that the
stochastic differential of $Y$ exists.
Then, calculating formally, we find
%
\renewcommand{\theequation}{\arabic{equation}}
\setcounter{equation}{32}
\begin{eqnarray}
\label{dY} \mathrm{d} Y_t &=& g(0+)\omega_{t-}\,\mathrm{d}L_t
+ \int_{-\infty}^t g'(t-s)
\omega_{s-}\,\mathrm{d}L_s \,\mathrm{d}t
+
q(0+)a_t \,\mathrm{d}t \nonumber
\\[-8pt]
\\[-8pt]
\nonumber
&&{} +\int_{-\infty}^tq'(t-s)a_s
\,\mathrm{d}s \,\mathrm{d}t,
\end{eqnarray}
which indicates that $Y$ can be represented, for $t \geq0$, as
%
\begin{equation}
Y_t = Y_0 + g(0+) \int_0^t
\omega_{s-}\,\mathrm{d}\overline{L}_s + \int_0^t
A_s \,\mathrm{d}s.
\end{equation}
%
%
Clearly, under the conditions formulated in Proposition~\ref{ThmSM},
the above integrals are well defined, and
$Y$, defined by (\ref{SMRep}), is a semimartingale, and $dY$ exists and
satisfies equation~(\ref{dY}).
A direct rewrite now shows that (\ref{dY}) agrees with the defining
equation (\ref{LSS}) of~$Y$, and
we can
then deduce that $Y$ is a semimartingale.
\end{pf*}

\begin{pf*}{Proof of Proposition~\ref{ThmQV}}
The result follows directly from the representation (\ref{SMRep}) and
from properties of the quadratic variation process, see, for example, Protter
\cite{Protter}.
\end{pf*}

\begin{pf*}{Proof of Proposition~\ref{propforward-generalambit}}
First, write
\[
\int_{-\infty}^TG(T,s)\omega_{s-} \,
\mathrm{d}L_s=\int_{-\infty
}^tG(T,s)
\omega_{s-} \,\mathrm{d}L_s+\int_t^TG(T,s)
\omega_{s-} \,\mathrm{d}L_s
\]
and observe that the first integral on the right-hand side is $\mathcal
{F}_t$-measurable. The result
follows by using double conditioning, first with respect to the $\sigma
$-algebra $\mathcal{G}_T$ generated by the paths of
$\omega_s, s\leq T$ and $\mathcal{F}_t$, and next with respect to
$\mathcal{F}_t$.
\end{pf*}

\begin{pf*}{Proof of Proposition~\ref{propforward-bmambit}}
By the Girsanov change of measure, we have
\begin{eqnarray*}
\int_{-\infty}^TG(T,s)\omega_{s-} \,
\mathrm{d}B_s &=& \int_{-\infty}^0G(T,s)
\omega_{s-} \,\mathrm{d}B_s + \int_{0}^TG(T,s)
\omega_{s-} \,\mathrm{d}B_s
\\
&=&\int_{0}^TG(T,s)\theta(s) \,\mathrm{d}s+\int
_{-\infty}^TG(T,s)\omega_{s-} \,
\mathrm{d}W_s ,
\end{eqnarray*}
where we set $B_s = W_s$ for $s< 0$.
By following the argumentation in the proof of Proposition \ref
{propforward-generalambit}, we are led to calculate
the expectation
\[
\E_{\eta} \biggl[\exp \biggl(\frac12\int_t^TG^2(T,s)
\omega^2_s \,\mathrm{d}s \biggr) \Big| \mathcal{F}_t
\biggr].
\]
But, by the stochastic Fubini theorem, see, for example, Barndorff-Nielsen and
Basse-O'Connor~\cite{BNBasse2009},
\begin{eqnarray*}
&&\int_t^T G^2(T,s)\int
_{-\infty}^s i(s,v) \,\mathrm{d}U_v \,
\mathrm{d}s
\\
&&\qquad=\int_t^T\int_{-\infty}^tG^2(T,s)i(s,v)
\,\mathrm{d}U_v \,\mathrm{d}s+\int_t^T
\int_t^sG^2(T,s)i(s,v) \,
\mathrm{d}U_v \,\mathrm{d}s
\\
&&\qquad=\int_{-\infty}^t\int_t^TG^2(T,s)i(s,v)
\,\mathrm{d}s \,\mathrm{d}U_v+\int_t^T
\int_v^TG^2(T,s)i(s,v) \,
\mathrm{d}s \,\mathrm{d}U_v .
\end{eqnarray*}
Using the adaptedness to $\mathcal{F}_t$ of the first integral and the
independence from $\mathcal{F}_t$
of the second, we find the desired result.
\end{pf*}

\begin{pf*}{Proof of Proposition~\ref{ProfForwardAffine}}
If $G(t,s)=g_1(t)g_2(s)$ it holds that
\[
\int_{-\infty}^TG(T,s)\omega_{s-} \,
\mathrm{d}W_s=\frac{g_1(T)}{g_1(t)}\int_{-\infty}^tG(t,s)
\omega_{s-} \,\mathrm{d}W_s =\frac{g_1(T)}{g_1(t)}\overline
Y_t .
\]
Similarly, if $i(t,s)=i_1(t)i_2(s)$,
\begin{eqnarray*}
\int_{-\infty}^t\int_t^TG^2(T,v)i(v,s)
\,\mathrm{d}v \,\mathrm{d}U_s&=&i_1^{-1}(t)\int
_t^TG^2(T,v)i_1(v) \,
\mathrm{d}v\int_{-\infty}^t i(t,s) \,
\mathrm{d}U_s
\\
&=& i_1^{-1}(t)\int_t^TG^2(T,v)i_1(v)
\,\mathrm{d}v Z_t ,
\end{eqnarray*}
and affinity holds in both the volatility and the spot price.

Opposite, to have affinity in $\overline Y_t$ we must have that
\[
\int_{-\infty}^tG(T,s)\omega_{s-} \,
\mathrm{d}W_s=\xi(T,t)\int_{-\infty
}^tG(t,s)
\omega_{s-} \,\mathrm{d}W_s
\]
for some function $\xi(T,t)$, which means that the ratio
$\xi(T,t)=G(T,s)/G(t,s)$ is independent of $s$. $\xi(T,t)$ is
differentiable in $T$ as long as $G$ is. Furthermore,
$\xi(T,T)=1$ by definition. Thus, by first differentiating $\xi$ with
respect to $T$ and next letting $T=t$, it holds that
\[
G_T(t,s)=\xi_T(t,t)G(t,s) ,
\]
where we use the notation $G_T=\partial G/\partial T$ and $\xi_T=\partial\xi/\partial T$ for the corresponding partial derivatives
with respect to the first argument.
Hence, we must have that
\[
G(t,s)=G(s,s)\exp \biggl(\int_s^t
\xi_T(u,u) \,\mathrm{d}u \biggr) ,
\]
and the separation property holds.

Likewise, to have affinity in the volatility $Z(t)$, we must have that
$\int_t^TG^2(T,v)i(v,s) \,\mathrm{d}v/\break i(t,s)$ must be independent of $s$. Denote
the ratio by
$\xi(T,t)$, and differentiate with respect to $T$ to obtain
\[
G^2(T,T)i(T,s)+2\int_t^TG(T,v)G_T(T,v)i(v,s)
\,\mathrm{d}v=\xi_T(T,t)i(t,s) .
\]
Hence,
\[
i(T,s)=-\int_t^TI(T,v)i(v,s) \,
\mathrm{d}v+J(T,t)i(t,s)
\]
for $I(T,t)=2G^{-2}(T,T)G(T,v)G_T(T,v)$ and $J(T,t)=G^{-2}(T,T)\xi_T(T,t)$. Differentiating with respect to
$T$, and next letting $T=t$ gives
\[
i_T(t,s)=i(t,s) \bigl(J_T(t,t)-I(t,t) \bigr) .
\]
Whence,
\[
i(t,s)=i(s,s)\exp \biggl(\int_s^t
\bigl(J_T(v,v)-I(v,v)\bigr) \,\mathrm{d}v \biggr) ,
\]
and the separation property holds for $i$. The proposition is proved.
\end{pf*}

\begin{pf*}{Proof of Proposition~\ref{PropRiskNeutralDynamicsF}}
Let
$H_T(t,s)=\int_t^TG^2(T,v)i(v,s) \,\mathrm{d}v$.
From Proposition~\ref{propforward-bmambit}, we have that
\[
F_t(T)=\Theta(t,T)\exp \biggl(\int_{-\infty}^tG(T,s)
\omega_{s-} \,\mathrm{d}W_s+\frac12\int
_{-\infty}^tH_T(t,s) \,
\mathrm{d}U_s \biggr)
\]
for a deterministic function $\Theta(t,T)$ given by
\begin{eqnarray*}
\Theta(t,T)=\Lambda(T)\exp \biggl(\int_{0}^TG(T,s)
\theta(s) \,\mathrm{d}s+\int_t^T
\phi^{\eta}_U \biggl(\frac{1}{2}H_T(v,v)
\biggr) \,\mathrm{d}v \biggr).
\end{eqnarray*}
Note that the process
$
M_T(t)\triangleq\int_{-\infty}^tG(T,s)\omega_{s-} \,\mathrm{d}W_s
$
is a (local) $Q^{\theta,\eta}$-martingale for $t\leq T$. Moreover,
from the stochastic Fubini theorem it holds that
\[
\int_{-\infty}^tH_T(t,s) \,
\mathrm{d}U_s=\int_{-\infty}^tH_T(s,s)
\,\mathrm{d}U_s+ \int_{-\infty}^t\int
_{-\infty}^u\frac{\partial H_T}{\partial u}(u,s) \,
\mathrm{d}U_s \,\mathrm{d}u ,
\]
where
we note that
$
\frac{\partial H_T}{\partial u}(u,s)=-G^2(T,u)i(u,s) .
$
Hence,
\begin{eqnarray*}
\int_{-\infty}^tH_T(t,s) \,
\mathrm{d}U_s &= \int_{-\infty}^tH_T(s,s)
\,\mathrm{d}U_s - \int_{-\infty}^t
G^2(T,u)\omega_u^2 \,\mathrm{d}u .
\end{eqnarray*}
The result is then a direct consequence of the It\^o formula for
semimartingales, see, for example, Protter~\cite{Protter}.
\end{pf*}
\begin{pf*}{Proof of Proposition~\ref{OptionPrice}}
From Proposition
\ref{propforward-bmambit}, we know that we can write the forward price as
\begin{eqnarray*}
F_{\tau}(T)&=& \Theta(\tau, T) \exp \biggl( \int_{-\infty}^{\tau}G(T,s)
\omega_{s-} \,\mathrm{d}W_s +\int_{-\infty}^{\tau}
\frac12 H_T(s,s) \,\mathrm{d}U_s \\
&&\hspace*{54pt}{}- \int
_{-\infty}^{\tau}\frac12 G^2(T,s)
\omega_s^2 \,\mathrm{d}s \biggr).
\end{eqnarray*}
Let now $p(x)=f(\exp(x))$, and suppose that $p\in L^1(\R)$. Recall that
$
p(x)=\frac1{2\pi}\int_{\R}\widehat{p}(y){\mathrm{e}}^{\mathrm
{i}xy} \,\mathrm{d}y$,
with $\widehat{p}(y)$ is the Fourier transform of $p(x)$ defined by
$
\widehat{p}(y)=\int_{\R}p(x){\mathrm{e}}^{-\mathrm{i}xy} \,\mathrm{d}x$.
Suppose that $\widehat p \in L^1(\mathbb{R})$.
Hence, we find
\begin{eqnarray*}
f\bigl(F_{\tau}(T)\bigr) &=& f\bigl(\exp\bigl(\ln\bigl(F_{\tau}(T)
\bigr)\bigr)\bigr) = p\bigl(\ln\bigl(F_{\tau}(T)\bigr)\bigr) =\frac1{2\pi}\int
_ {\R}\widehat{p}(y){\mathrm{e}}^{\mathrm
{i}y\ln(F_{\tau
}(T))}\,\mathrm{d}y
\\
&=& \frac1{2\pi}\int_ {\R}\widehat{p}(y){
\mathrm{e}}^{\mathrm{i}y\ln
\Theta(\tau
,T)}\exp\biggl(\mathrm{i}y\biggl(\int_{-\infty}^{\tau}G(T,s)
\omega_{s-} \,\mathrm{d}W_s+\int_{-\infty}^{\tau}
\frac12 H_T(s,s) \,\mathrm{d}U_s\\
&&\hspace*{108pt}\qquad{}- \frac12\int_{-\infty}^{\tau}G^2(T,s)
\omega_s^2\,\mathrm{d}s \biggr)\biggr) \,\mathrm{d}y .
\end{eqnarray*}
Next, by commuting integration and expectation using dominated
convergence and $\mathcal{F}_t$-adaptedness, we obtain
\begin{eqnarray*}
C_t &=&{\mathrm{e}}^{-r(\tau-t)}\frac1{2\pi}\int
_{\R}\widehat{p}(y) \exp \bigl(\mathrm {i}y \ln\Theta(\tau,T)
\bigr)
\\
&&\hspace*{59pt}{} \times\exp \biggl(\mathrm{i}y \biggl(\int_{-\infty}^tG(T,s)
\omega_{s-} \,\mathrm{d}W_s+\int_{-\infty}^t
\frac12 H_T(s,s) \,\mathrm{d}U_s \\
&&\hspace*{86pt}\qquad{}- \int
_{-\infty}^t \frac12 G^2(T,s)
\omega_s^2\,\mathrm{d}s \biggr) \biggr)
\\
&&\hspace*{59pt}{} \times\E_{Q} \biggl[\exp \biggl( \mathrm{i}y \biggl(\int
_t^{\tau}G(T,s)\omega_{s-}\,
\mathrm{d}W_s + \int_{t}^{\tau} \frac12
H_T(s,s)\,\mathrm{d}U_s \\
&&\hspace*{106pt}\qquad{}- \int_t^{\tau}
\frac12 G^2(T,s)\omega_s^2\,\mathrm{d}s
\biggr) \biggr)\Big | \mathcal{F}_t \biggr] \,\mathrm{d}y ,
\end{eqnarray*}
which holds by the stochastic Fubini theorem. Using the independent
increment property of $U$ and double conditioning, we reach
\begin{eqnarray*}
A&:=&\E_{Q} \biggl[\exp \biggl( \mathrm{i}y \biggl(\int_t^{\tau}G(T,s)
\omega_{s-}\,\mathrm{d}W_s + \int_{t}^{\tau}
\frac12 H_T(s,s)\,\mathrm{d}U_s - \int
_t^{\tau} \frac12 G^2(T,s)
\omega_s^2\,\mathrm{d}s \biggr) \biggr)\Big |
\mathcal{F}_t \biggr]
\\
&=&\E_{\eta} \biggl[\exp \biggl(\mathrm{i}y \int_{t}^{\tau}
\frac12 H_T(s,s)\,\mathrm{d}U_s-\frac12 y^2
\int_t^{\tau}G^2(T,s)
\omega_{s}^2 \,\mathrm{d}s - \mathrm{i}y \int_t^{\tau}
\frac12 G^2(T,s)\omega_s^2\,\mathrm{d}s
\biggr) \Big| \mathcal{F}_t \biggr]
\\
&=& \E_{\eta} \biggl[\exp \biggl(\mathrm{i}y \int_{t}^{\tau}
\frac12 H_T(s,s)\,\mathrm{d}U_s+a \int
_t^{\tau}G^2(T,s)\int
_{-\infty}^s i(s,v)\,\mathrm{d}U_v \,
\mathrm{d}s \biggr) \Big| \mathcal{F}_t \biggr],
\end{eqnarray*}
where
we define $a:=a(y):= -\frac12 y^2-\frac12 \mathrm{i} y$.
Using the stochastic Fubini theorem again, we get
\begin{eqnarray*}
A &=& \E_{\eta} \biggl[\exp \biggl(\mathrm{i}y\int_{t}^{\tau}
\frac12 H_T(s,s)\,\mathrm{d}U_s+a \int
_{-\infty}^{\tau}\int_{v}^{\tau}
G^2(T,s) i(s,v) \,\mathrm{d}s \,\mathrm{d}U_v \biggr)\Big |
\mathcal{F}_t \biggr]
\\
&=& \E_{\eta}\biggl[\exp\biggl(\int_{t}^{\tau}
\mathrm{i}y\frac12 H_T(s,s)\,\mathrm{d}U_s
\\
&&\hspace*{36pt}{}+a \int_{-\infty}^{t}\int_{v}^{\tau}
G^2(T,s) i(s,v) \,\mathrm{d}s \,\mathrm{d}U_v +a \int
_{t}^{\tau}\int_{v}^{\tau}
G^2(T,s) i(s,v) \,\mathrm{d}s \,\mathrm{d}U_v\biggr) \Big|
\mathcal{F}_t\biggr]
\\
&=& \exp \biggl( a \int_{-\infty}^{t}\int
_{v}^{\tau} G^2(T,s) i(s,v) \,
\mathrm{d}s \,\mathrm{d}U_v \biggr)
\\
&&{}\times \E_{\eta} \biggl[\exp \biggl(\mathrm{i}y \int_{t}^{\tau}
\frac12 H_T(s,s)\,\mathrm{d}U_s +a \int
_{t}^{\tau}\int_{v}^{\tau}
G^2(T,s) i(s,v) \,\mathrm{d}s \,\mathrm{d}U_v \biggr)\Big |
\mathcal{F}_t \biggr]
\\
&= &\exp \biggl( a \int_{-\infty}^{t}\int
_{v}^{\tau} G^2(T,s) i(s,v) \,
\mathrm{d}s \,\mathrm{d}U_v \biggr)
\\
& &{}\times\E_{\eta} \biggl[\exp \biggl(\int_{t}^{\tau}
\biggl\{ \mathrm{i}y \frac12 H_T(v,v) +a \int_{v}^{\tau}
G^2(T,s) i(s,v) \,\mathrm{d}s \biggr\} \,\mathrm{d}U_v
\biggr) \Big| \mathcal{F}_t \biggr].
\end{eqnarray*}
Altogether, we obtain
\begin{eqnarray*}
C_t&=&{\mathrm{e}}^{-r(\tau-t)}\frac1{2\pi}\int
_{\R}\widehat {p}(y)\exp \biggl(\mathrm {i}y \biggl(\ln
\Theta(\tau,T)+\int_{-\infty}^tG(T,s)
\omega_{s-} \,\mathrm{d}W_s\\
&&\hspace*{97pt}\qquad{}-\int_{-\infty}^t
\frac{1}{2}G^2(T,s)\omega_s^2\,
\mathrm{d}s \biggr) \biggr)
\\
&&\hspace*{60pt}{} \times\exp \biggl(\int_{-\infty}^t \biggl\{ \mathrm{i} y
\frac12 H_T(v,v)+a\int_v^{\tau}G^2(T,s)i(s,v)
\,\mathrm{d}s \biggr\} \,\mathrm{d}U_v \biggr)
\\
&&\hspace*{60pt}{} \times\exp \biggl(\int_t^{\tau}
\phi_U^{\eta} \biggl(\mathrm{i}y \frac12 H_T(v,v)+a\int
_v^{\tau}G^2(T,s)i(s,v) \,\mathrm{d}s
\biggr) \,\mathrm{d}v \biggr) \,\mathrm{d}y .
\end{eqnarray*}
The above expression can be further simplified by noting that
\begin{eqnarray*}
&&\mathrm{i} y \frac12 H_T(v,v)+a\int_v^{\tau}G^2(T,s)i(s,v)
\,\mathrm{d}s
\\
&&\qquad=-\frac12 y^2\int_v^{\tau}G^2(T,s)i(s,v)
\,\mathrm{d}s + \mathrm{i}y \frac12 \biggl(\int_v^TG^2(T,s)i(s,v)
\,\mathrm{d}s-\int_v^{\tau}G^2(T,s)i(s,v)
\,\mathrm{d}s \biggr)
\\
&&\qquad=-\frac12 y^2\int_v^{\tau}G^2(T,s)i(s,v)
\,\mathrm{d}s + \mathrm{i}y \frac12 \int_{\tau
}^TG^2(T,s)i(s,v)
\,\mathrm{d}s =: h(T, \tau, v, y).
\end{eqnarray*}
Then
\begin{eqnarray*}
C_t&=&{\mathrm{e}}^{-r(\tau-t)}\\
&&{}\times\frac1{2\pi}\int
_{\R}\widehat {p}(y)\exp \biggl(\mathrm {i}y \biggl(\ln
\Theta(\tau,T)+\int_{-\infty}^tG(T,s)
\omega_{s-} \,\mathrm{d}W_s-\int_{-\infty}^t
\frac{1}{2}G^2(T,s)\omega_s^2\,
\mathrm{d}s \biggr) \biggr)
\\
& &\hspace*{39pt}{}\times\exp \biggl(\int_{-\infty}^t h(T, \tau, v,
y) \,\mathrm{d}U_v \biggr) \exp \biggl(\int_t^{\tau}
\phi_U^{\eta} \bigl(h(T, \tau, v, y)\bigr) \,\mathrm{d}v
\biggr) \,\mathrm{d}y .
\end{eqnarray*}
\upqed\end{pf*}

\begin{pf*}{Proof of Proposition~\ref{propforward-arithmetic}}
Observe that
\[
\E_{\theta,\eta} \biggl[\int_{-\infty}^TG(T,s)
\omega_{s-} \,\mathrm{d}L_s \Big| \mathcal{F}_t
\biggr]=(-\mathrm{i})\frac{\mathrm{d}}{\mathrm{d}x} \E_{\theta,\eta} \biggl[\exp \biggl(
\mathrm{i}x\int_{-\infty
}^TG(T,s)\omega_{s-}
\,\mathrm{d}L_s \biggr) \Big| \mathcal{F}_t
\biggr]_{x=0} .
\]
We then proceed as in the proof of Proposition \ref
{propforward-generalambit}, and finally we perform the differentiation
and let $x=0$.
\end{pf*}

\begin{pf*}{Proof of Proposition~\ref{PropCorGammaKernel}}
We have
\begin{eqnarray*}
g^2(t)=\frac{\newlambda^{2\newnu-1}}{\Gamma(2\newnu-1)} t^{(2\newnu
-1)-1}\exp (-\newlambda t )=
\ga(t;2\newnu-1,\newlambda) = q(t),
\end{eqnarray*}
which is a probability density and hence
$\int_0^{\infty}g^2(t)\,\mathrm{d}t = 1$.
Now we derive the explicit formula for the autocorrelation function.
\begin{eqnarray*}
\int_0^{\infty}g(t+h)g(t)\,\mathrm{d}t =
\frac{\newlambda^{2\newnu-1}}{\Gamma(2\newnu-1)} \exp \biggl(-\frac{\newlambda h}{2} \biggr) \int
_0^{\infty} \bigl(t(t+h)\bigr)^{\newnu-1}\exp (-
\newlambda t )\,\mathrm{d}t.
\end{eqnarray*}
Note that according to Gradshteyn and Ryzhik~\cite{GrRy2007}, Formula 3.383.8
\begin{eqnarray*}
\int_0^{\infty} \bigl(t(t+h)\bigr)^{\newnu-1}
\exp (-\newlambda t )\,\mathrm{d}t = \frac{1}{\sqrt{\pi}} \biggl(\frac{h}{\newlambda}
\biggr)^{\newnu
-
{1}/{2}}\exp \biggl({\frac{\newlambda h}{2}} \biggr)\Gamma(\newnu
)K_{
{1}/{2}-\newnu} \biggl(\frac{\newlambda h}{2} \biggr)
\end{eqnarray*}
for $|\operatorname{arg}(h)|< \pi$ and $\operatorname{Re}(\newlambda),
\operatorname{Re}(\newnu)>0$,
where $K_{\newnu}$ is the modified Bessel function of the third kind.
Hence,
\begin{eqnarray*}
\int_0^{\infty}g(t+h)g(t)\,\mathrm{d}t &=&
\frac{\newlambda^{2\newnu-1}}{\Gamma(2\newnu-1)} \exp \biggl(-\frac{\newlambda h}{2} \biggr) \frac{1}{\sqrt{\pi
}}
\biggl(\frac
{h}{\newlambda} \biggr)^{\newnu-{1}/{2}}\exp \biggl({\frac
{\newlambda
h}{2}}
\biggr)\Gamma(\newnu)K_{{1}/{2}-\newnu} \biggl(\frac
{\newlambda
h}{2} \biggr)
\\
&=&\frac{(\newlambda h)^{\newnu-1/2}}{\Gamma(2\newnu-1)} \frac{1}{\sqrt{\pi}} \Gamma(\newnu)K_{{1}/{2}-\newnu} \biggl(
\frac{\newlambda
h}{2} \biggr).
\end{eqnarray*}
Now we apply Gradshteyn and Ryzhik~\cite{GrRy2007}, Formula 8.335.1,
to obtain
\begin{eqnarray*}
\Gamma(2\newnu-1)= \frac{2^{2\newnu-2}}{\sqrt{\pi}}\Gamma \biggl(\newnu -\frac{1}{2}
\biggr)\Gamma(\newnu).
\end{eqnarray*}
Then
\begin{eqnarray*}
\int_0^{\infty}g(t+h)g(t)\,\mathrm{d}t %
&=&\frac{(\newlambda h)^{\newnu-1/2}}{({2^{2\newnu-2}}/{\sqrt {\pi }})\Gamma (\newnu-{1}/{2} )\Gamma(\newnu)}
 \frac{1}{\sqrt{\pi}} \Gamma(\newnu)K_{{1}/{2}-\newnu} \biggl(
\frac{\newlambda
h}{2} \biggr)
\\
&=&\frac{(\newlambda h)^{\newnu-1/2}}{2^{2\newnu-2}\Gamma
(\newnu
-{1}/{2} )} K_{{1}/{2}-\newnu} \biggl(\frac{\newlambda h}{2} \biggr)\\
& =&
\frac{(\newlambda h)^{\newnu-1/2}}{2^{\newnu-{1}/{2}}2^{\newnu
-{3}/{2}}\Gamma (\newnu-{1}/{2} )} K_{{1}/{2}-\newnu} \biggl(\frac{\newlambda h}{2} \biggr).
\end{eqnarray*}
Since $K_{\newnu}(x)=K_{-\newnu}(x)$ according to
Gradshteyn and Ryzhik~\cite{GrRy2007}, Formula 8.486.16, the result follows.
\end{pf*}
\end{appendix}

\section*{Acknowledgements}
We would like to thank Andreas Basse-O'Connor and Jan Pedersen for helpful
discussions and constructive comments.
Also, we are a grateful to the valuable comments by two anonymous
referees and by the Editor.
F.E. Benth is grateful for the financial support from the project
``Energy Markets: Modelling, Optimization and Simulation (EMMOS)''
funded by the Norwegian Research Council under grant eVita/205328.
Financial support by the Center for Research in Econometric
Analysis of Time Series, CREATES, funded by the Danish National
Research Foundation is gratefully
acknowledged by A.E.D. Veraart.



\printhistory

\end{document}